\documentclass[11pt, letter]{article}
\usepackage{amsmath, amsthm, amssymb}
\usepackage[left=1in, bottom=1in, right=1in, top=1in]{geometry}
\usepackage{cancel} 
\usepackage{xcolor}
\usepackage{orcidlink}
\usepackage{amsmath,amssymb,amsthm}

\newtheorem{proposition}{Proposition}[section]

\usepackage[comma,authoryear,round]{natbib}
\usepackage{soul,ulem}
\usepackage{graphicx}
\usepackage{booktabs}
\usepackage{dsfont}

\usepackage{float}
\usepackage{bbm}
\usepackage{authblk}
\usepackage{comment}
\usepackage{longtable}
\usepackage[font=normalsize]{caption}  
\usepackage[title]{appendix}
\usepackage{csquotes}
\usepackage{caption}
\usepackage{lipsum}
\usepackage{hyperref}

\title{\textcolor{black}{GPU-accelerated Regularized Vector Multiplicative Error Models for Positive-valued Financial Time Series}}
\author[1]{Rohan Hemant Chhatre \orcidlink{0009-0007-1200-0832}\thanks{E-mail: rohan.chhatre@uconn.edu (Corresponding author)}}
\author[1]{Chiranjit Dutta \orcidlink{0009-0009-7803-3308}\thanks{E-mail: chiranjituconn1@gmail.com}}
\author[2]{Sumanta Basu \orcidlink{0000-0003-0433-7359}\thanks{E-mail: sumbose@cornell.edu}}
\author[1]{Nalini Ravishanker \orcidlink{0000-0002-2028-4771}\thanks{E-mail: nalini.ravishanker@uconn.edu}}
\affil[1]{{Department of Statistics, University of Connecticut, Storrs, CT 06269, USA}}
\affil[2]{{Department of Statistics and Data Science, Cornell University, Ithaca, NY 14853, USA}}
\date{}
\begin{document}
\maketitle
\begin{abstract}
{\color{black}
The logarithmic vector multiplicative error model (log-vMEM) provides a methodology for modeling and forecasting multivariate positive-valued 
time series. 
However, the number of autoregressive parameters increases rapidly with the system dimension and lag order, posing challenges for estimation and model selection. This paper develops a regularized estimation framework for the log-vMEM with multivariate Gamma errors by incorporating hierarchical lag structures to induce sparsity and accommodate heterogeneous lag dynamics. We consider three hierarchical lag structures: componentwise, own-other, and elementwise, in combination with group lasso, adaptive group lasso, group MCP, and group SCAD penalties. Parameter estimation is carried out through an alternating block-coordinate optimization algorithm with sequential Gauss-Seidel updates, while GPU-accelerated numerical quadrature is employed to address the computational burden of evaluating the multivariate Gamma likelihood. Simulation experiments investigate parameter recovery and lag selection under different sparsity structures. The real data analysis is carried out on the realized volatility measures for Microsoft (\texttt{NASDAQ:MSFT}) which allows a comparison of the competing regularization approaches in an empirical financial setting. Together, these developments provide a computational framework for sparse multivariate volatility modeling with flexible lag structures and dependent Gamma errors.
}
\end{abstract} 
\vspace{5pt}

\section{Introduction} \label{intro}
We study the problem of modeling and forecasting multivariate positive valued financial time series. Positive-valued processes are ubiquitous in financial time series representing several measures related to high-frequency trading activity such as 
realized volatilities, absolute returns, daily ranges, trading volumes, durations, market depth, etc. They 
share similar stylized features such as non-negativity, persistence, and clustering patterns. The dynamic structure of the Multiplicative Error Model (MEM), introduced by \cite{engle2002new} (please see \cite{cipollini2022multiplicative}, for a survey) as a generalization of the Autoregressive Conditional Duration (ACD) model \citep{engle1998autoregressive}, is designed to capture 
stylized 
features of such time series. 
The positive-valued variable of interest is modeled as the product of its conditional expectation and an i.i.d. error process with unit mean, with the conditional expectation  
itself being modeled as a linear combination of its own past lagged value(s) and lagged values of the observed variable. One of the early applications of univariate MEM 
is the the conditional autoregressive range (CARR) model for modeling the daily range \citep{chou2005forecasting}.

The  
vector Multiplicative Error Model (vMEM) 
provides a useful mechanism for analyzing multivariate positive-valued variables \citep{manganelli2005duration,cipollini2006vector,hautsch2011econometrics,cipollini2013semiparametric}. The form of the vMEM model for $\boldsymbol{x}_t$ is 

\begin{align} 
    \boldsymbol{x}_t & = \boldsymbol{\mu}_t \odot \boldsymbol{\epsilon}_t, \quad \boldsymbol{\epsilon}_t \sim \text{D}^{+}(\mathbf{1},\boldsymbol{\Sigma}), \label{VMEM1} \\
    \boldsymbol{\mu}_t & = \boldsymbol{\delta} + \sum_{i=1}^{p}  \boldsymbol{E}^{(i)} \boldsymbol{x}_{t-i} 
    + \sum_{j=1}^{q} \boldsymbol{F}^{(j)} \boldsymbol{\mu}_{t-j},
    \label{VMEM2}
\end{align}
where $\boldsymbol{\mu}_t$ is the conditional mean of $\boldsymbol{x}_t$ given its history, the term $\boldsymbol{\mu}_t \odot \boldsymbol{\epsilon}_t$ represents the  
Hadamard product of $\boldsymbol{\mu}_t$ and vector-valued errors  $\boldsymbol{\epsilon}_t$. The errors $\boldsymbol{\epsilon}_t$ are assumed to follow a 
%
%
multivariate positive-valued distribution, $\text{D}^{+}$,  with mean $\mathbf{1}$ (vector of ones), and variance-covariance matrix $\boldsymbol{\Sigma}$, which is assumed to be positive definite. 
In \eqref{VMEM2}, $\boldsymbol{\mu}_t$ is 
modeled as a constant $d$-dimensional intercept 
$\boldsymbol{\delta}$ plus a 
linear combination of its own past lagged values,  $\boldsymbol{\mu}_{t-j},~j=1,\ldots,q$, and lagged values of the observed vector, i.e., 
$\boldsymbol{x}_{t-i},~i=1,\ldots,p$,
with $\boldsymbol{E}^{(i)},~i=1,\dots,p$ 
and $\boldsymbol{F}^{(j)},~j=1,\dots,q$
being $d \times d$ parameter matrices. 
Note that $\boldsymbol{\mu}_{t}$ will be non-negative for all $t$, 
if 
the elements in $\boldsymbol{\delta}$ 
are positive, and all the elements in $\boldsymbol{E}^{(i)},~i=1,\dots,p$ and $\boldsymbol{F}^{(j)},~j=1,\dots,q$ are non-negative \citep{he2004extended}. 
Several choices 
have been proposed in literature for the distribution of the errors $\boldsymbol{\epsilon}_t$, 
including the Gamma-copula distribution 
\citep{cipollini2017copula}, 
and the bivariate Gamma  distribution
\citep{ahoniemi2009joint}. 
In these papers, the  authors obtained maximum likelihood estimates of the vMEM parameters.
A semi-parametric vMEM with 
the generalized method of moments (GMM) approach was described in 
\cite{cipollini2013semiparametric}, 
while \cite{donelli2021bayesian} used a Bayesian semi-parametric approach to model the error vector as a Dirichlet Process Mixture of multivariate log-normal distributions. 
Irrespective of the estimation approach, positivity of the estimated parameters must be verified in order to ensure positivity of $\boldsymbol{\mu}_t$, which is not required if we instead model $\log \boldsymbol{\mu}_t$.



The logarithmic vector MEM  (log-vMEM) 
\citep{taylor2017logarithmic} has the form
\begin{align}
    \boldsymbol{x}_t & = \boldsymbol{\mu}_t \odot \boldsymbol{\epsilon}_t, \quad \boldsymbol{\epsilon}_t \sim \text{D}^{+}(\mathbf{1},\boldsymbol{\Sigma}), \label{logVMEM1} \\
    \log \boldsymbol{\mu}_t & = \boldsymbol{\omega} + \sum_{i=1}^{p}  \boldsymbol{A}^{(i)} \log \boldsymbol{x}_{t-i} 
    + \sum_{j=1}^{q} \boldsymbol{B}^{(j)} \log \boldsymbol{\mu}_{t-j}. \label{logVMEM2}
\end{align}
The model specification for $\boldsymbol{x}_t$ in \eqref{logVMEM1} is the same as the specification in \eqref{VMEM1}, while
\eqref{logVMEM2} models the {logarithm} of the conditional mean $\boldsymbol{\mu}_t$ as a 
linear combination of 
$\log \boldsymbol{x}_{t-i},~i=1,\dots,p$ and 
$\log \boldsymbol{\mu}_{t-j},~j=1,\dots,q$.  
Here,
$\boldsymbol{\omega}$ is a $d$-dimensional vector of constants, and $\boldsymbol{A}^{(i)},~i=1,\dots,p$ and $\boldsymbol{B}^{(j)},~j=1,\dots,q$ are $d \times d$ parameter matrices.
\cite{taylor2017logarithmic}
assumed multivariate lognormal errors
in their log-vMEM specification to model the joint dynamics of trading duration, volume and volatility of NYSE stocks, while \cite{cipollini2017copula} assumed Gamma-copula distribution of the errors in their vMEM specification to model the interdependence of realized volatility, volume and number of trades of Johnson \& Johnson stock (\texttt{NASDAQ: JNJ}).

{\color{black}
Our contributions in this article are fourfold. First, we specify a correction to the multivariate gamma (MVG) specification \citep{mathai1991multivariate,tsionas2004bayesian} for the error vector in the log-vMEM model in~\eqref{logVMEM1}. This specification provides a framework for jointly modeling positive-valued multivariate observations while accounting for the dependence between the components of the error vector.

\textcolor{black}{Second, we incorporate componentwise, own-other, and elementwise hierarchical lag structures \citep{nicholson2017varx,nicholson2020high} with convex and non-convex group penalties, enabling structured sparsity and heterogeneous lag selection within a multivariate Gamma likelihood-based estimation framework.
The HLag framework incorporates lag-order selection directly into the regularization problem, thereby addressing parameter 
growth and the 
\textit{curse of dimensionality} while preserving the underlying lag structure of the model.}

\textcolor{black}{Third, we develop an alternating block coordinate optimization procedure for both unpenalized and penalized likelihood estimation, using Gauss–Seidel updates and derivative-free optimization of the parameter blocks.}

Finally, the fourth contribution of this study is the adoption of the GPU-accelerated adaptive numerical integration procedure to evaluate the multivariate gamma likelihood. The proposed variant of the likelihood function substantially reduces the computational burden associated with the numerical integration step and improves the scalability of the model to higher-dimensional settings.
We use our GPU-accelerated approach for modeling the joint dynamics of intraday volatility measures of the Microsoft asset(\texttt{NASDAQ: MSFT}). The model performance measures are studied in a rolling window moving origin style validation technique over 27 possible origins.
}

The format of the paper follows: Section \ref{Section:log-vMEM} describes 
the log-likelihood function for the log-vMEM with multivariate Gamma errors. The computation and maximization of this log-likelihood is performed using the Block Coordinate Descent algorithm which is stated in Section~\ref{sec:maxloglike}. {\color{black}The results of a simulation study are presented in Section~\ref{secsimstudy} demonstrating the parameter recovery by the proposed unpenalized model}. Section \ref{sec:pmle} describes HLag based penalty structures and the penalized maximum likelihood estimation procedure. Section \ref{section:simulated_data_analysis} presents simulated data analysis demonstrating true active structures could be recovered using the proposed method.
Section~\ref{emp_anal} presents an analysis of the multivariate intraday measures of volatility for Microsoft (\texttt{NASDAQ: MSFT})
{\color{black}
for the various competing models using the relatively slower CPU version for one validation set. 
Section~\ref{sec:gpu} provides a GPU-accelerated framework of the data analysis for the competing models implemented across 27 rolling window validation sets. 
Finally, Section \ref{section:concl} concludes with a discussion and directions for future research.
The \texttt{R} code for our analysis is available on the \href{https://github.com/Rohan-Chhatre/sparse_logvMEM}{GitHub link}.
}
\section{Logarithmic Vector Multiplicative Error Models} \label{Section:log-vMEM}

Let $\boldsymbol{x}_t,~t=1,\ldots T$ be a $d$-dimensional positive-valued time series with $\boldsymbol{x}_t = (x_{t1},\ldots, x_{td})'$. Let 
$\boldsymbol{\mu}_t = E[\boldsymbol{x}_t|\mathcal{F}_{t-1}]$ denote the conditional mean process, where $\mathcal{F}_{t-1}$ is the information set available up to time $t-1$. Let $\boldsymbol{\epsilon}_t = (\epsilon_{t1},\ldots,\epsilon_{td})'$ be the error vector at time $t$ whose support is $\mathbb{R}^{+d}$.

Setting $q=0$ in \eqref{logVMEM2} 
gives the log-vMEM$(p)$ specification
\begin{align} \label{logVMEMp}
    \boldsymbol{x}_t & = \boldsymbol{\mu}_t \odot \boldsymbol{\epsilon}_t, \quad \boldsymbol{\epsilon}_t \sim \text{D}^{+}(\mathbf{1},\boldsymbol{\Sigma}), \notag \\
    \log \boldsymbol{\mu}_t & = \boldsymbol{\omega} + \sum_{i=1}^{p}  \boldsymbol{A}^{(i)} \log \boldsymbol{x}_{t-i}, 
\end{align}
where $\odot$ represents the Hadamard elementwise product, $\text{D}^{+}$ denotes 
a multivariate positive-valued distribution with mean $\mathbf{1}$ (vector of ones), and variance-covariance matrix $\boldsymbol{\Sigma}$, 
$\boldsymbol{\omega}$ is a $d$-dimensional real-valued vector, and $\boldsymbol{A}^{(i)}$ 
are $d \times d$ parameter matrices, for $i=1,\ldots,p$.
The first two conditional moments of $\boldsymbol{x}_t$ are 
\begin{align}
\text{E}(\boldsymbol{x}_t|\mathcal{F}_{t-1}) & = \boldsymbol{\mu}_t, \notag \\
\text{Cov}(\boldsymbol{x}_t|\mathcal{F}_{t-1}) & = \boldsymbol{\mu}_t \boldsymbol{\mu}'_{t} \odot \boldsymbol{\Sigma} = \mathbf{V}_t, \text{ say}.
\label{eq:xmoments}
\end{align}


{
\color{black}
For the purpose of this study, we use the multivariate gamma p.d.f. 
in~(4) from \citet{tsionas2004bayesian}. Consider the random vector $\mathbf{X} = [X_1,X_2,\ldots X_d]$ having the 
\begin{equation*}    
X_i=V_i+Z,\qquad i=1,\ldots,d,
\label{eq:tsionas_rv}
\end{equation*}
where $V_1,\ldots,V_d$ are independent $\operatorname{Gamma(\lambda_i,\theta_i); \ (i=1,2\ldots d)}$, and $Z$ $\sim$ $\operatorname{Gamma(\alpha,\beta)}$. The multivariate gamma p.d.f. is given as 
\begin{equation}
\begin{aligned}
f_T(\mathbf{x};\boldsymbol{\Theta})
&=
\frac{\beta^\alpha}{\Gamma(\alpha)}
\prod_{i=1}^{d}
\frac{\theta_i^{\lambda_i}}{\Gamma(\lambda_i)}
\prod_{i=1}^{d}
\left\{
\int_0^{B}
(x_i-z)^{\lambda_i-1}
e^{-\theta_i(x_i-z)}
z^{\alpha-1}e^{-\beta z}\,dz
\right\},
\end{aligned}
\label{eq:tsionas_density}
\end{equation}
where $B=\min_{1\leq i\leq d}x_i$,
$\mathbf{x}\in(0,\infty)^d$, and all parameters, $\boldsymbol{\Theta} = (\alpha,\beta,\boldsymbol{\xi},\boldsymbol{\theta})$ are positive.
We propose a correction to this p.d.f since \eqref{eq:tsionas_density} does not integrate to one (shown in Appendix~\ref{app:normalization_issue}). The corrected p.d.f $g(\cdot;\boldsymbol{\Theta})$ has been shown to integrate to one in Appendix~\ref{app:corrected_density}, and the p.d.f can be represented as
\begin{equation}
\begin{aligned}
g(\mathbf{x};\boldsymbol{\Theta})
&=
\frac{\beta^\alpha}{\Gamma(\alpha)}
\prod_{i=1}^d
\frac{\theta_i^{\lambda_i}}
{\Gamma(\lambda_i)}
\times
\int_0^B
\left\{
\prod_{i=1}^d
(x_i-z)^{\lambda_i-1}
e^{-\theta_i(x_i-z)}
\right\}
z^{\alpha-1}e^{-\beta z}\,dz
\end{aligned}
\label{eq:corrected_mvg_density}
\end{equation}
where $B=\min_{1\leq i\leq d}x_i$,
$\mathbf{x}\in(0,\infty)^d$, and all parameters are positive. We adopt the corrected
density as the error distribution in the log-vMEM model, assuming
$\boldsymbol{\epsilon}_t \sim g(\mathbf{x};\boldsymbol{\Theta})
$
with p.d.f.
}
\textcolor{black}{
\begin{equation} \label{Eq:MVG}   g(\boldsymbol{\epsilon}_t) = \frac{\beta^{\alpha}}{\Gamma(\alpha)} \prod_{i=1}^{d} \frac{\theta_i^{\xi_i}}{\Gamma(\xi_i)} \int_{0}^{B_t} 
\left\{\prod_{i=1}^{d} (\epsilon_{ti} - Z_t)^{(\xi_i - 1)}  \exp[-\theta_i(\epsilon_{ti} - Z_t)] Z_t^{\alpha - 1} \exp(-\beta Z_t) dZ_t\right\}, \quad \color{black}\epsilon_{ti} > Z_{t}>0
\end{equation}
where $B_t = \text{min}\{\epsilon_{ti}: i = 1,\ldots,d\}$, $t=1,\ldots,T$ and $Z_t \sim$ Gamma$(\alpha,\beta)$ with p.d.f., i.e., 
$$
f(z_{t}) = \frac{\beta^{\alpha}}{\Gamma(\alpha)} z_{t}^{\alpha - 1} \exp(-\beta z_{t}), z_{t}>0, \beta >0,
$$ 
with shape parameter $\alpha$ and rate $\beta$. Let $\boldsymbol{\xi} = (\xi_1,\ldots,\xi_d)'$ and $\boldsymbol{\theta} = (\theta_1,\ldots,\theta_d)', \theta_i >0, \ \xi_i>0$, for $i=1,\ldots,d$. 
For $t=1,\ldots,T$, the theoretical moments of $\boldsymbol{\epsilon}_{t}$ are
\begin{align}
    \text{E}(\epsilon_{ti})  &= \dfrac{\xi_{i}}{\theta_{i}} + \dfrac{\alpha}{\beta}, \quad i=1,\ldots,d,  \text{ and} \notag \\
     \text{Cov}(\epsilon_{ti},\epsilon_{tj})  & = \dfrac{\xi_{i}}{\theta_{i}^2} \mathbf{1}_{ij}   
    + \dfrac{\alpha}{\beta^2}\quad i,j=1,\ldots,d,
    \label{eq:epsmoments}
\end{align}
where $\mathbf{1}_{ij}=1$ if $i=j$, and zero otherwise.
We set 
\begin{align}
\alpha=\beta/2, \ \xi_{i} = \beta/2,  \text{ and } \theta_{i}=\beta,~ i = 1,\ldots,d,
\label{setparams}
\end{align}
to ensure that $\text{E}(\boldsymbol{\epsilon}_{t}) = \mathbf{1}$, while
Cov$(\boldsymbol{\epsilon}_{t})$ is an equi-correlated covariance matrix whose diagonal elements are $1/\beta$ and off-diagonal elements are $1/(2\beta)$.
\cite{taylor2017logarithmic} discussed the model in \eqref{logVMEM1} \& \eqref{logVMEM2} with multivariate lognormal errors having a unit mean vector in order to model the joint dynamics of trading duration, volume, and volatility. The corresponding correlation matrix has unit diagonal elements and constant off-diagonal elements equal to $1/2$. The specification provides a parsimonious way of incorporating contemporaneous dependence without introducing additional pairwise covariance parameters, which is particularly useful as the dimension of the system increases.
}

\subsection{Log-likelihood function} \label{subsection:loglik}
 
Let $\boldsymbol{X}= (\boldsymbol{x}_{1}',\ldots,\boldsymbol{x}_{T}')'$ be the $T \times d$ observed data matrix,
 where $\boldsymbol{x}_{t} = (x_{t1},\ldots,x_{td})',~t= 1,\ldots,T$.
Let $\boldsymbol{x}_t \oslash \boldsymbol{\mu}_t = \big(\frac{x_{t1}}{\mu_{t1}},\ldots,\frac{x_{td}}{\mu_{td}}\big)'$. 
The conditional p.d.f. of $\boldsymbol{x}_t$ given the history $\mathcal{F}_{t-1}$ at time $t$ is
\begin{align}
    f(\boldsymbol{x}_t|\mathcal{F}_{t-1}) = g(\boldsymbol{x}_t \oslash \boldsymbol{\mu}_t) \bigg |\frac{\partial{\boldsymbol{\epsilon}_t}}{\partial{\boldsymbol{x}_t}}\bigg|
\end{align}
where $\oslash$ represents the Hadamard elementwise division, and $g(.)$ is the p.d.f. shown in 
\eqref{Eq:MVG}.
Let $\boldsymbol{\Theta} = (\boldsymbol{\omega},\beta,\text{vec}'(\boldsymbol{A}^{(1)}),\ldots, \text{vec}'(\boldsymbol{A}^{(p)}))'$ denote the 
vector of all parameters (of dimension 
$d^2p+d+1$), where vec is the vectorization operator.
The conditional likelihood function of 
$\Theta$ given the data is
\begin{align}
        L(\Theta; \boldsymbol{X}) &= \prod_{t=p + 1}^{T} f(\boldsymbol{x}_t|\mathcal{F}_{t-1}; \boldsymbol{\Theta}) \notag \\
   & = \bigg( \prod_{t = p+1}^{T} \prod_{i = 1}^{d} \dfrac{1}{\mu_{ti}}\bigg) \times \bigg(\prod_{t=p+1}^{T} g(\boldsymbol{x}_{t} \oslash \boldsymbol{\mu}_{t})\bigg) 
   \label{likelihood}
\end{align}
and the conditional log-likelihood is 
\begin{align} \label{Eq:log-likelihood}
    \ell(\Theta;\boldsymbol{X}) & = (T-p) (d+1) \bigg(\frac{\beta}{2} \log \beta - \log\Gamma\Big(\frac{\beta}{2}\Big)\bigg) - \sum_{t=p+1}^{T} \sum_{i=1}^{d} \log(\mu_{ti}) \notag \\ 
    & - \beta \sum_{t=p+1}^{T} \sum_{i=1}^{d} \frac{x_{ti}}{\mu_{ti}} + \sum_{t=p+1}^{T} \log \Bigg(\int_{0}^{B_{t}} \bigg\{\prod_{i=1}^{d} \bigg(\frac{x_{ti}}{\mu_{ti}}- Z_{t}\bigg)^{\big(\frac{\beta}{2}-1\big)}\bigg\}  
    \times Z_{t}^{\big(\frac{\beta}{2}  - 1 \big)} \times {\color{black}e^{\beta(d-1)Z_t}} \times  dZ_{t}\Bigg)
\end{align}
where the dependence on $\boldsymbol{\omega}$ and $\text{vec}(\boldsymbol{A}^{(1)}),\ldots, \text{vec}(\boldsymbol{A}^{(p)}))$ 
is through the conditional mean dynamics in \eqref{logVMEMp}, i.e.,
\begin{align*}
    \log \boldsymbol{\mu}_t  = \boldsymbol{\omega} + \sum_{i=1}^{p}  \boldsymbol{A}^{(i)} \log \boldsymbol{x}_{t-i} 
\end{align*}
{\color{black}
To reduce the dimensionality of the numerical optimization problem, we use the stationary mean relationship of the log-vMEM process to express the intercept vector $\boldsymbol{\omega}$ in terms of the autoregressive coefficient matrices $\boldsymbol{A}^{(1)},\ldots,\boldsymbol{A}^{(p)}$ and the multivariate Gamma parameter $\beta$. Since the unconditional logarithmic mean of the observed process is unknown, we subsequently approximate it using a second-order Taylor expansion and replace the corresponding population moments with their sample estimates. This yields a plug-in parameterization of the intercept and reduces the number of independently optimized parameters from $d^2p+d+1$ to $d^2p+1$.
The resulting conditional log-likelihood is evaluated as a
function of
$\boldsymbol{\Theta} =
(\beta,\operatorname{vec}'(\boldsymbol{A}^{(1)}),\ldots,
\operatorname{vec}'(\boldsymbol{A}^{(p)}))'$.}
\begin{align}
& \bigg(\boldsymbol{\mathbf{I}} - \sum_{i=1}^{p} \boldsymbol{A}^{(i)}\bigg) \text{E}(\log \boldsymbol{x}_{t})  = \boldsymbol{\omega} + \text{E}(\log \boldsymbol{\epsilon}_{t}), \rm{ i.e.}, \notag \\
 & \boldsymbol{\omega}   =  \bigg(\boldsymbol{\mathbf{I}} - \sum_{i=1}^{p} \boldsymbol{A}^{(i)}\bigg) \text{E}(\log \boldsymbol{x}_{t}) - \text{E}(\log \boldsymbol{\epsilon}_{t}).
    \label{omega_derivation}
\end{align}
{\color{black}
We assume that the log-vMEM process is stationary with finite unconditional first and second moments, denoted by
$\mathrm{E}(\boldsymbol{x}_t)=\boldsymbol{\mu}
=(\mu_1,\ldots,\mu_d)'$ and
$\mathrm{Cov}(\boldsymbol{x}_t)=\boldsymbol{V}
=\{V_{ij}\}$. Since the logarithmic  specification in \eqref{logVMEMp} can be represented as a non-Gaussian VAR$(p)$ process for $\log\boldsymbol{x}_t$, we impose the usual stability condition that the roots of
\[
\left|\boldsymbol{I}
-\boldsymbol{A}^{(1)}z-\cdots
-\boldsymbol{A}^{(p)}z^p\right|=0
\]
lie outside the unit circle. This condition ensures stability of the logarithmic autoregressive dynamics under the assumed innovation process.}


From \eqref{setparams}, $\boldsymbol{\epsilon}_{t} \sim \text{MVG}(\beta/2,\beta,\boldsymbol{\beta}/2,\boldsymbol{\beta})$, where $\boldsymbol{\beta} = \beta \mathbf{1}$, so that
$\epsilon_{ti} \sim \text{Gamma}(\beta,\beta)$ \citep{mathai1991multivariate}. Using the fact that if $Y \sim \text{Gamma}(\alpha^{*},\beta^{*})$, then $\text{E}(\log Y) = \psi(\alpha^{*}) - \log (\beta^{*})$, where $\psi(.)$ is the digamma function, we see that 
{\color{black}
\begin{equation}\label{Eq:Approx_equation1}
    \mathrm{E}(\log\boldsymbol{\epsilon}_t)
    =
    \left[\psi(\beta)-\log(\beta)\right]\boldsymbol{1}_d,
\end{equation}}

Since the second order Taylor expansion around $E(Y)$ gives 
\begin{equation*} \label{Eq:Approx_equation2}
    \text{E}(\log Y) \approx \log(\text{E}(Y)) - \dfrac{\text{Var}(Y)}{2(\text{E}(Y))^2}, 
\end{equation*}
we get
\begin{equation}    
\text{E}(\log \boldsymbol{x}_{t}) \approx \log \boldsymbol{\mu} - \text{diag}(\boldsymbol{V}) \oslash 2\boldsymbol{\mu}^{\circ 2}
\label{eq:Elogxt}
\end{equation}

where $\text{diag}(\boldsymbol{V}) = (V_{11},\ldots, V_{dd})'$, $\oslash$ represents the Hadamard elementwise division and $\circ$ denotes the Hadamard elementwise power. 
\textcolor{black}{The approximation is applied componentwise and depends only on the marginal  unconditional means and variances of the observed process, and does not include off-diagonal elements of $\boldsymbol{V}$.}

{\color{black}
Following a moment-based approximation similar to \citet{cattivelli2020adaptive}, we replace the unknown unconditional mean $\boldsymbol{\mu}$ by its sample counterpart
$\hat{\boldsymbol{\mu}} =
T^{-1}\sum_{t=1}^{T}\boldsymbol{x}_t$, and the marginal variances $V_{ii}$ by
\[
\hat{V}_{ii}
=
\frac{1}{T-1}
\sum_{t=1}^{T}(x_{ti}-\bar{x}_i)^2,
\qquad i=1,\ldots,d,
\]
where $\bar{x}_i=T^{-1}\sum_{t=1}^{T}x_{ti}$. 
Combining these sample moments with the exact Gamma log-moment in \eqref{Eq:Approx_equation1} and the second-order approximation in \eqref{Eq:Approx_equation2} provides a plug-in estimate of the intercept vector for any given values of the autoregressive coefficients and $\beta$.
Substituting this expression into the conditional log-likelihood in \eqref{Eq:log-likelihood} yields the following reduced-dimensional plug-in conditional log-likelihood:
}


\begin{align} \label{Eq:simplog-likelihood}   \ell(\boldsymbol{\Theta};\boldsymbol{X}) & = (T-p) (d+1) \bigg(\frac{\beta}{2} \log \beta - \log\Gamma\Big(\frac{\beta}{2}\Big)\bigg) - \sum_{t=p+1}^{T} \sum_{i=1}^{d} \log(\mu_{ti}) \notag \\ 
    & - \beta \sum_{t=p+1}^{T} \sum_{i=1}^{d} \frac{x_{ti}}{\mu_{ti}} + \sum_{t=p+1}^{T} \log \Bigg(\int_{0}^{B_{t}}\bigg\{ \prod_{i=1}^{d} \bigg(\frac{x_{ti}}{\mu_{ti}}- Z_{t}\bigg)^{\big(\frac{\beta}{2}-1\big)}\bigg\}  
    \times Z_{t}^{\big(\frac{\beta}{2}  - 1 \big)}\times {\color{black}e^{\beta(d-1)Z_t}} \times dZ_{t}\Bigg),
\end{align}
where the dependence on $\beta$ and $\text{vec}(\boldsymbol{A}^{(1)}),\ldots, \text{vec}(\boldsymbol{A}^{(p)}))$ 
is through the conditional mean dynamics i.e.,
\begin{align*}
    \log \boldsymbol{\mu}_t  = \bigg(\boldsymbol{\mathbf{I}} - \sum_{i=1}^{p} \boldsymbol{A}^{(i)}\bigg) \bigg(\log \hat{\boldsymbol{\mu}} - \text{diag}(\boldsymbol{\hat{V}}) \oslash 2\boldsymbol{\hat{\mu}}^{\circ 2} \bigg) - (\psi(\beta) - \log(\beta))\mathbf{1}' + \sum_{i=1}^{p}  \boldsymbol{A}^{(i)} \log \boldsymbol{x}_{t-i} 
\end{align*}
{\color{black}

The approximation in \eqref{eq:Elogxt} provides a moment-based representation of $\mathbb{E}(\log \mathbf{x}_t)$ in terms of
the unconditional mean and marginal variances of
the observed process. This allows the intercept
vector $\boldsymbol{\omega}$ to be expressed in
terms of the autoregressive coefficients, the
Gamma parameter $\beta$, and the sample moments. This allows us to avoid separate numerical optimization over the intercept parameters.

An alternative would be to estimate
$\mathbb{E}(\log \mathbf{x}_t)$ directly using the sample
mean of the log-transformed observations. We retain the second-order moment approximation in the present study. 
The resulting estimator is therefore
a plug-in likelihood estimator rather than an
unrestricted conditional maximum likelihood
estimator.

This parameterization reduces the dimension of the optimization problem by $d$ and avoids a separate numerical optimization over the intercept vector. The approximation
does not alter the assumed multivariate Gamma 
density.
The numerical estimation procedure and its parameter recovery properties are examined in the following Section~\ref{secsimstudy},~\ref{section:simulated_data_analysis}.}

\subsection{Block Coordinate Descent  Maximization of the Log-likelihood Function}\label{sec:maxloglike}

\textcolor{black}{
The log-likelihood function $\ell(\boldsymbol{\Theta}; \boldsymbol{X})$ in \eqref{Eq:simplog-likelihood} is a non-convex function of $\boldsymbol{\Theta}$ and \textcolor{black}{cannot be maximized in closed form.}
We estimate the parameter vector $\boldsymbol{\Theta}$ by maximizing the log-likelihood function $\ell(\boldsymbol{{\Theta}};\boldsymbol{X})$ in \eqref{Eq:simplog-likelihood} using an alternating block coordinate descent algorithm \citep{bertsekas1999nonlinear}. This approach is adopted because direct joint maximization of the full likelihood equation is computationally arduous primarily due to the non-linearity and high-dimensionality of the parameter space. 
We therefore adopt a block co-ordinate optimization strategy \citep{tseng2001convergence,bertsekas1999nonlinear} that alternately updates the autoregressive coefficient matrix $\boldsymbol{A}$ and the multivariate gamma shape parameter $\beta$. Within each block update, the objective function is optimized using the derivative-free optimizer \texttt{bobyqa} proposed by \citet{powell2009bobyqa}.
}

\textcolor{black}{
Using good initial values in a non-linear equation solver leads to faster convergence. Initial parameter values for log-vMEM can be based on fitting log-ACD model \citep{bauwens2000logarithmic} to each $x_{ti}, i=1,\ldots,d$, since log-vMEM is the multivariate extension of log-ACD model. We assume $x_{ti}$ follows the log-ACD($p,0$) model as 
\begin{align}
    x_{ti} & = \mu_{ti} \epsilon_{ti} \notag \\
    \log \mu_{ti} & = \omega_{i} + \sum_{s=1}^{p} \alpha_{si} \log \epsilon_{t-s,i}
\end{align}
where $\epsilon_{ti} \sim Gamma(1,\beta_{i})$ with mean $1/\beta_{i}$ and variance $1/\beta_{i}^2$. We have used {R-package} \texttt{ACDm::acdFit()} \citep{belfrage2015acdm} to fit the log-ACD model to each component $x_{ti}$, $i= 1,\ldots, d$ and obtain the parameter estimates $\hat{\alpha}_{si}, \hat{\omega_{i}} \text{ and } \hat{\beta}_{i}$, where $s= 1$,$\ldots,p$; $i=1, \ldots, d$. We initialize
${\beta^{(0)}}
=
\operatorname{median}\left(\widehat{\beta}_1,\ldots,\widehat{\beta}_d\right)$,
and
$\boldsymbol{A^{(0)}}
=
\operatorname{diag}
\left(
\widehat{\alpha}_{11},\ldots,\widehat{\alpha}_{p1},
\ldots,
\widehat{\alpha}_{1d},\ldots,\widehat{\alpha}_{pd}
\right)$.
}

\texttt{nloptr::bobyqa()} performs derivative-free optimization by constructing a local quadratic approximation to the objective function using function evaluations at a set of interpolation points. For a generic parameter block $\boldsymbol{\theta}$, where $\boldsymbol{\theta}$ represents either a row $\boldsymbol{A}_i$ of $\boldsymbol{A}$ or the scalar parameter $\beta$, define the minimization objective as
$f(\boldsymbol{\theta})
=
-\ell(\boldsymbol{\theta};\mathbf{X})$.
At the current \texttt{nloptr::bobyqa()} iterate $\boldsymbol{\theta}^{(k)}$, a quadratic interpolation model of the objective function is constructed as
\begin{align}
\label{eq:eq:bobyqa_quad_interpolation}
q_k(\mathbf{s})
=
f\!\left(\boldsymbol{\theta}^{(k)}\right)
+
\mathbf{g}_k^{\prime}\mathbf{s}
+
\frac{1}{2}\mathbf{s}^{\prime}\mathbf{H}_k\mathbf{s},
\end{align}

where $\mathbf{s}$ denotes a trial step, and $\mathbf{g}_k$ and $\mathbf{H}_k$ are coefficients of the quadratic interpolation model and not the analytically solved gradient and Hessian values. We obtain a trial step by approximately solving the bound-constrained trust-region problem
\[
\mathbf{s}^{(k)}
=
\arg\min_{\mathbf{s}}
q_k(\mathbf{s}),
\qquad
\|\mathbf{s}\| \leq \Delta_k,
\qquad
\mathbf{u_1}
\leq
\boldsymbol{\theta}^{(k)}+\mathbf{s}
\leq
\mathbf{u_2},
\]
where $\Delta_k$ denotes the current trust-region radius, and $\mathbf{u_1}$ and $\mathbf{u_2}$ denote the lower and upper bounds on the parameters, respectively. The objective function is then evaluated at the proposed point
$\boldsymbol{\theta}^{(k)}+\mathbf{s}^{(k)}$, and this new function evaluation is used to update the quadratic interpolation model and the trust region. Thus, \texttt{bobyqa} does not require analytical derivatives of the log-likelihood; instead, it successively refines a local quadratic model using repeated evaluations of $\ell(\boldsymbol{\theta};\mathbf{X})$ until its convergence criterion is satisfied. \citet{powell2009bobyqa} describes more details about \texttt{nloptr::bobyqa()}.


\textcolor{black}{
Generating several sets of starting values, we 
run the \texttt{bobyqa} function from each set and select the set of values and the corresponding log-likelihood function 
with the largest (maximum) log-likelihood.
The  log-likelihood maximization implements a scheme with two loops. 
In the inner loop, each row-wise (or scalar for $\beta$) sub-problem is solved by the derivative-free local optimizer \texttt{bobyqa}. The outer loop is the Gauss-Seidel {block coordinate descent (BCD) algorithm}, which is discussed below.
Note that for convergence, the classical Gauss–Seidel BCD algorithm assume that each block subproblem is solved  exactly. In using the \texttt{nloptr::bobyqa()} function, we get an {approximate} row-wise optimum rather than a global optimum.
}

\textcolor{black}{
Starting from \(\boldsymbol{\Theta_0^{(0)}}=({\beta^{(0)}},\operatorname{vec}(\boldsymbol{A^{(0)}})')'\), the BCD algorithm iterates for \(j=0,1,\ldots\):
\begin{align}    
\boldsymbol{A}^{(j+1)}
=
\arg\max_{\boldsymbol{A}}
\ell\!\left(\boldsymbol{A},{\beta^{(j)}};\mathbf X\right),
\label{eq:A_update}
\end{align}
where the optimization over \(\boldsymbol{A}\) is performed row-by-row, updating each row sequentially while holding all remaining rows fixed.
Next, holding \(\boldsymbol{A}^{(j+1)}\) fixed, we update
\begin{align}
{\beta}^{(j+1)}
=
\arg\max_{{\beta}}
\ell\!\left(\boldsymbol{A}^{(j+1)},{\beta};\mathbf X\right).
\label{eq:beta_update}
\end{align}
The row-wise updates of \(\boldsymbol{A}\) are performed using \texttt{bobyqa}, implemented in the R package \texttt{nloptr} \citep{johnson2024nloptr}, while the scalar optimization for \({\beta}\) is carried out using one-dimensional numerical optimization. The row-wise updates of the coefficient matrix \({\boldsymbol{A}}\) are performed sequentially using a Gauss-Seidel block coordinate descent scheme, where each updated row is immediately incorporated into subsequent row updates within the same sweep; please see \citet{wright2015coordinate} for additional details.
}

The convergence criterion is based on the relative
change in the log-likelihood after each complete
outer iteration of the alternating block coordinate
optimization algorithm. Let
$\ell^{(j)}=\ell(\boldsymbol{A}^{(j)},\beta^{(j)};\mathbf{X})$
denote the log-likelihood evaluated at iteration
$j$. We define the relative change as
\begin{align}
\Delta_{\ell}^{(j)}
=
\frac{
\left|
\ell^{(j+1)}-\ell^{(j)}
\right|
}{
\max\left\{1,\left|\ell^{(j)}\right|\right\}
}
\label{eq:del_A}
\end{align}
To avoid termination due to a single small change
in the log-likelihood, we require the convergence
criterion to be satisfied for consecutive outer
iterations. Let $c^{(0)}=0$ and define
\begin{equation}
c^{(j+1)}
=
\begin{cases}
c^{(j)}+1,
& \text{if } \Delta_{\ell}^{(j)}
<\varepsilon_{\ell},\\
0,
& \text{otherwise},
\end{cases}
\end{equation}
where $\varepsilon_{\ell}=10^{-5}$ is tolerance.

The algorithm terminates when $c^{(j+1)}\geq 2$,
or when the maximum number of outer iterations
is reached. 
This provides a
numerical stopping criterion based on the
 stabilization of the log-likelihood.

Note that evaluating the likelihood and maximizing it are not separable steps here, they are nested. The \texttt{bobyqa} function behaves like any numerical optimizer, and repeatedly evaluates the log-likelihood (objective function) at trial points, and then uses the evaluations to build its local model and propose the next point. So within each blockwise update of the 
algorithm, the \texttt{bobyqa} calls the log-likelihood function, which itself requires the numerical integral over $Z_t$ many times per row, searching over  $\mathbf{A}_i \in \mathbb{R}^{1 \times dp}$,  until convergence. 



We report CPU times for a single evaluation of the log-likelihood function. We vary $d$ from 5 to 500, and $T$ from 100 to 4500, with $p=10$ held constant. These results are summarized in Table~\ref{CPU_time_by_T_d}. These simulations were carried out on a personal laptop equipped with an Apple MacBook Pro (14-inch, Nov 2024) equipped with an Apple M4 Pro chip, 24GB of unified memory, and running a 64-bit version of macOS.


\begin{table}[H]
\centering
\begin{tabular}{|c|cccccc|}
\hline
 
& $\boldsymbol{d=5}$
& $\boldsymbol{d=10}$
& $\boldsymbol{d=50}$
& $\boldsymbol{d=100}$
& $\boldsymbol{d=200}$
& $\boldsymbol{d=500}$ \\
\hline

\textbf{T=100}
& 0.01
& 0.01
& 0.01
& 0.04
& 0.10
& 0.62 \\

\textbf{T=500}
& 0.05
& 0.05
& 0.07
& 0.18
& 0.57
& 3.35 \\

\textbf{T=1000}
& 0.08
& 0.09
& 0.17
& 0.36
& 1.10
& 6.55 \\

\textbf{T=1500}
& 0.12
& 0.12
& 0.24
& 0.51
& 1.61
& 9.81 \\

\textbf{T=3000}
& 0.26
& 0.26
& 0.48
& 1.05
& 3.19
& 18.98 \\

\textbf{T=4500}
& 0.38
& 0.39
& 0.73
& 1.60
& 4.87
& 29.08 \\

\hline
\end{tabular}
\caption{CPU times (in seconds) for a single evaluation of the log-likelihood function for different sample sizes ($T$) and dimensions ($d$). The log-likelihood compute time increases sharply with sample size($T$) as well as dimensions($d$).}
\label{CPU_time_by_T_d}
\end{table}

\subsection{Results from a Simulation Study} \label{secsimstudy}

We present a simulation study to show that the parameters of log-vMEM with multivariate gamma errors are estimated accurately using our approach. We generate $B=100$ sets (replicates) of data from log-vMEM in \eqref{logVMEMp},\eqref{Eq:MVG}, setting $d=5$, $p=1$ and $T=1500$. The true values are set as $\beta = 10$, $\boldsymbol{\omega} = (0.1,0.1,0.1,0.1,0.1)'$, and  
$$\boldsymbol{A} = 
\begin{bmatrix}
-0.052 & 0.377 & 0.131 & 0 & 0\\
0 & 0.106 & 0 & 0.122 & 0\\
0 & 0.113 & 0.056 & 0.022 & 0\\
0 & 0 & 0.266 & 0.041 & 0.365\\
0.357 & 0 & 0 & 0.854 & 0.077
\end{bmatrix}.
$$ 
Note that in order to ensure stationarity of $\boldsymbol{x}_{t}$, we have followed the procedure in \citet{lutkepohl2005new, wilms2021sparse} to select the true values of the coefficient matrix $\boldsymbol{A}$. 
\begin{table}[H]


\begin{center}
\begin{tabular}{|c|c|c|c|c|}
\hline
\textbf{Parameter} &\textbf{ True value} &\textbf{Average Estimate} & \textbf{Average Bias} & \textbf{Average RMSE} \\ 
\hline
$\beta$ & 10.000 & 10.011 & -0.011 & 0.189 \\ 
  $A_{11}$ & -0.052 & -0.056 & 0.004 & 0.026 \\ 
  $A_{21}$ & 0.000 & -0.004 & 0.004 & 0.026 \\ 
  $A_{31}$ & 0.000 & -0.004 & 0.004 & 0.026 \\ 
  $A_{41}$ & 0.000 & -0.001 & 0.001 & 0.029 \\ 
  $A_{51}$ & 0.357 & 0.358 & -0.001 & 0.026 \\ 
  $A_{12}$ & 0.377 & 0.379 & -0.002 & 0.029 \\ 
  $A_{22}$ & 0.106 & 0.103 & 0.003 & 0.028 \\ 
  $A_{32}$ & 0.113 & 0.112 & 0.001 & 0.029 \\ 
  $A_{42}$ & 0.000 & -0.001 & 0.001 & 0.031 \\ 
  $A_{52}$ & 0.000 & 0.001 & -0.001 & 0.030 \\ 
  $A_{13}$ & 0.131 & 0.129 & 0.002 & 0.031 \\ 
  $A_{23}$ & 0.000 & 0.002 & -0.002 & 0.029 \\ 
  $A_{33}$ & 0.056 & 0.058 & -0.002 & 0.028 \\ 
  $A_{43}$ & 0.266 & 0.265 & 0.001 & 0.029 \\ 
  $A_{53}$ & 0.000 & -0.000 & 0.000 & 0.031 \\ 
  $A_{14}$ & 0.000 & -0.006 & 0.006 & 0.026 \\ 
  $A_{24}$ & 0.122 & 0.114 & 0.008 & 0.026 \\ 
  $A_{34}$ & 0.022 & 0.018 & 0.004 & 0.023 \\ 
  $A_{44}$ & 0.041 & 0.033 & 0.008 & 0.027 \\ 
  $A_{54}$ & 0.854 & 0.845 & 0.009 & 0.027 \\ 
  $A_{15}$ & 0.000 & -0.004 & 0.004 & 0.019 \\ 
  $A_{25}$ & 0.000 & -0.001 & 0.001 & 0.017 \\ 
  $A_{35}$ & 0.000 & -0.003 & 0.003 & 0.018 \\ 
  $A_{45}$ & 0.365 & 0.363 & 0.002 & 0.017 \\ 
  $A_{55}$ & 0.077 & 0.075 & 0.002 & 0.017 \\ 
   \hline
\end{tabular}
\end{center}
\caption{Simulation results for the logarithmic vector multiplicative error model with multivariate Gamma errors based on $B=100$ replications with sample size $T=1500$, $d=5$, and $p=1$. For each parameter, the table reports the true value, and the average estimate, bias, and RMSE averaged across the replications. 
\textcolor{black}{The table reports robust parameter recovery with relatively small $\text{RMSE}$ and $\text{Bias}$.}
The code for the simulation runs for higher dimensions and a variety of sample sizes have been made available on \href{https://github.com/Rohan-Chhatre/sparse_logvMEM}{GitHub}.}
\label{MLE_logvMEM_simulation}
\end{table}

\textcolor{black}{
Let $\theta$ be generic notation for any parameter and $\hat{\theta}_b$ denote its estimate under replication $b$. We report the following summary metrics across replications. 
\begin{align}
& \text{Average Estimate}  = \frac{1}{B} \sum_{b=1}^{B}\hat{\theta}_{b}, \\ \nonumber
& \text{Average Bias} = \frac{1}{B} \sum_{b=1}^{B}(\theta - \hat{\theta}_{b}), \\ \nonumber
&\text{Average RMSE} = \sqrt{\frac{1}{B} \sum_{b=1}^{B}(\theta - \hat{\theta}_{b})^2}.
\end{align}
}
\textcolor{black}{
Table \ref{MLE_logvMEM_simulation} provides these summary statistics for all the parameters.
The average estimates closely approximate the respective true values for all the parameters, and exhibit low average bias and low average RMSE. 
}

\section{Penalized Maximum Likelihood Estimation}\label{sec:pmle}

We discuss 
estimation of the log-vMEM$(p)$ parameters by maximizing
the simplified log-likelihood function 
$\ell({\boldsymbol{\Theta}};\boldsymbol{X})$ in 
\eqref{Eq:simplog-likelihood}
under different sparsity inducing penalties, which are reviewed in Section~\ref{section:HLag}.

\subsection{Hierarchical lag (HLag) based penalty structures }\label{section:HLag}

Following \cite{nicholson2020high}, we employ three hierarchical lag based penalty structures, i.e., componentwise, elementwise, and own-other, using their notation as well. 


For $1\leq \ell \leq {\color{black}p}$, let
$\mathbf{A}^{(\ell)},~\ell=1,\ldots,p$ be $d \times d$ coefficient matrices, 
let $\mathbf{A} = [\mathbf{A}^{(1)},\ldots,\mathbf{A}^{(p)}] \in \mathbb{R}^{d \times dp}$, and 
$\mathbf{A}^{(\ell:p)} = [\mathbf{A}^{(\ell)},\ldots,\mathbf{A}^{(p)}] \in \mathbb{R}^{d \times d(p-\ell+1)}$.
Let $\mathbf{A}_{i}$ be the $i^{th}$ row of $\mathbf{A}$ and
    $\mathbf{A}_{ij}^{(a)}$ be the $(i,j)^{th}$ element of the matrix $\mathbf{A}^{(a)}$, so that 
    $\mathbf{A}_{ij}^{(\ell:p)} = [\mathbf{A}_{ij}^{(\ell)},\ldots,\mathbf{A}_{ij}^{(p)}] \in \mathbb{R}^{1 \times (p-\ell+1)}$ is a $(p-\ell+1)$-dimensional row vector.
    Let $\mathbf{A}_{i}^{(a)}$ be the $i^{th}$ row of $\mathbf{A}^{(a)}$ and 
    $\mathbf{A}_i^{(\ell:p)} = [\mathbf{A}_{i}^{(\ell)},\ldots,\mathbf{A}_{i}^{(p)}] \in \mathbb{R}^{1 \times d(p-\ell+1)}$ is a $d(p-\ell+1)$-dimensional row vector.
Let $\mathbf{L} \in \mathbb{R}^{d \times d}$ denote  the maximal coefficient lag (maxlag) matrix, and let $\mathbf{L}_{ij} = \text{ max} \{\ell : \mathbf{A}_{ij}^{(\ell)} \neq 0\}$
be the $(i,j)^{th}$ element of $\mathbf{L}$, so that 
$\mathbf{L}_{ij}=0$ if $\mathbf{A}_{ij}^{(\ell)} = 0$ for all $\ell = 1,\ldots,p$. That is,  $\mathbf{L}_{ij}$ is the smallest $\ell$ such that $\mathbf{A}_{ij}^{([\ell+1]:p)} = \boldsymbol{0}$. Let $L_{i}$ be a positive integer denoting lags for $i=1,\ldots d$.

The \textbf{Componentwise (C)} sparse structure allows each of the $d$ marginal equations from \eqref{Eq:simplog-likelihood} to have its own maxlag, while requiring all components within each equation to share the same maxlag, i.e.,
        $\mathbf{L}_{ij} = L_{i},~\forall \ j$, for $i=1,\ldots,d$.
    This implies $\mathbf{A}_{i}^{([L_{i} + 1]:p)} = \boldsymbol{0}$ and $\mathbf{A}_{ij}^{(L_{i})} \neq 0$, for some $j$.

The \textbf{Own-Other (O)} sparse structure is similar to the componentwise structure, but includes  an additional within-lag hierarchy which imposes an assumption that the own lag of a series is more informative than lags of other series, that is, each component series has longer temporal dependence than cross dependencies, i.e., 
        $\mathbf{L}_{ij} = L_{i}^{\text{other}}, \quad \text{for } i \neq j,\text{ and } \mathbf{L}_{ii} \in \{L_{i}^{\text{other}},L_{i}^{\text{other}}+1\},\quad \text{for} \quad i=1,\ldots,d$.

The \textbf{Elementwise (E)} sparsity structure  is the most flexible structure with no predefined relationships among the elements of $\mathbf{L}$.

\begin{figure}[H]
    \centering
    \includegraphics[scale=0.3]{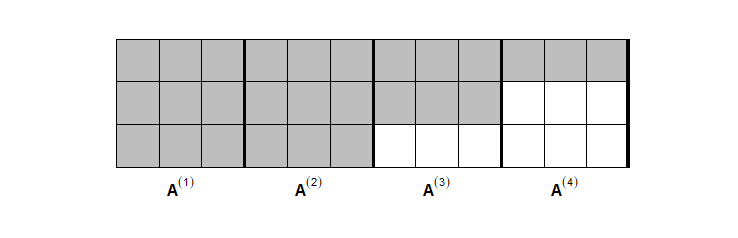}
    \includegraphics[scale=0.3]{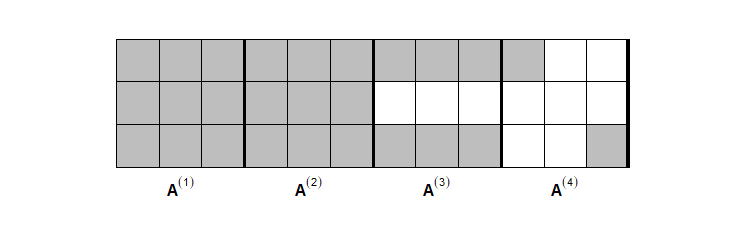}
    \includegraphics[scale=0.3]{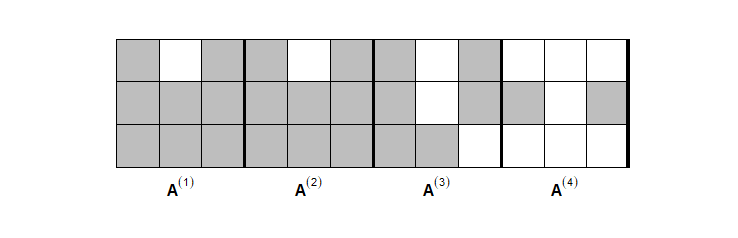}
    \caption{Illustration of Componentwise, Own other, and Elementwise HLag active set structure (shaded in grey)}
    \label{fig:componentwise}
\end{figure}

We illustrate below the maxlag matrix for the three situations when 
$p =4$
and $d=3$, while Figure~\ref{fig:componentwise} shows the corresponding active set structures (for more details, see \cite{nicholson2020high}):
\begin{eqnarray*}
\mathbf{L}^{C} = \begin{bmatrix}
4 & 4 & 4\\
3 & 3 & 3\\
2 & 2 & 2
\end{bmatrix}, \ 
 \mathbf{L}^{O} = \begin{bmatrix}
                4 & 3 & 3\\
                2 & 2 & 2\\
                3 & 3 & 4
            \end{bmatrix}, \ 
\mathbf{L}^{E} = \begin{bmatrix}
                3 & 0 & 3\\
                4 & 2 & 4\\
                3 & 3 & 2
            \end{bmatrix}.
               \end{eqnarray*}



We employ the group lasso and adaptive group lasso penalties \citep{yan2017hierarchical,lim2015learning,zhao2009composite}, 
as well as    
non-convex group penalties such as the group SCAD (smoothly clipped absolute deviation) penalty \citep{fan2001variable} and the group MCP (minimax concave penalty) \citep{breheny2015group}. 
Nested groups lead to hierarchical sparsity constraints where zeroing out one set of parameters implies that the other set is also zero. The forms of these penalties are shown below.




The minimax concave penalty function (MCP) \citep{zhang2010nearly} is defined as
\begin{equation}
    P^{MCP}_{\lambda,\gamma}(\theta) = 
    \begin{cases}
        \lambda \theta - \dfrac{\theta^2}{2\gamma} & \text{if } \theta \leq \gamma \lambda\\
        \dfrac{1}{2}\gamma \lambda^2 & \text{if } \theta > \gamma \lambda.
    \end{cases}
\end{equation}
We have used $\gamma =3$ for MCP as in \cite{breheny2015group}, while $\lambda$ is the regularization parameter. As in \cite{fan2001variable}, the SCAD penalty is defined as
\begin{equation}
    P^{SCAD}_{\lambda,\gamma}(\theta) = 
    \begin{cases}
        \lambda \theta & \text{if } \theta \leq \lambda\\
        \dfrac{\gamma \lambda \theta - 0.5 (\theta^2 + \lambda^2)}{\gamma - 1} & \text{if } \lambda < \theta \leq \gamma \lambda \\ 
        \dfrac{\lambda^2(\gamma + 1)}{2} & \text{if } \theta > \gamma \lambda. 
    \end{cases}
\end{equation}
Following \cite{fan2001variable}, we let $\gamma = 3.7$, while $\lambda$ is the regularization parameter.

For the adaptive group lasso, let $\widetilde{\boldsymbol{A}}$ denote
the unpenalized estimates of the autoregressive coefficient matrices. The adaptive weights are constructed using these initial estimates, with an exponent of one and a small threshold $\delta=0.005$ to avoid excessively large weights.

For the componentwise structure, define
\begin{equation*}
w_{i,\ell}^{C}
=
\left[
\max\left\{
\|\widetilde{\mathbf{A}}_i^{(\ell:p)}\|_2,
\delta
\right\}
\right]^{-1}.
\end{equation*}

For the elementwise and own-other structures, let
$\widetilde{\boldsymbol{A}}^{\mathrm{clip}}$ denote the
coefficient wise transformation
\[
\widetilde{\mathbf{A}}_{ij}^{\mathrm{clip},(\ell)}
=
\max\{\widetilde{\mathbf{A}}_{ij}^{(\ell)},\delta\}.
\]
The corresponding adaptive weights are
\begin{align*}
w_{ij,\ell}^{E}
&=
\left[
\|\widetilde{\mathbf{A}}_{ij}^{\mathrm{clip},(\ell:p)}\|_2
\right]^{-1},\\
w_{i,\ell}^{O}
&=
\left[
\max\left\{
\left\|
\left(
\widetilde{\mathbf{A}}_{i,-i}^{\mathrm{clip},(\ell)},
\widetilde{\mathbf{A}}_i^{\mathrm{clip},([\ell+1]:p)}
\right)
\right\|_2,
\delta
\right\}
\right]^{-1}
\end{align*}
Here, $w_{i,\ell}^{O}$ is defined for
$\ell=1,\ldots,p-1$. The weights are computed from
the unpenalized estimates and remain fixed during
penalized optimization.

The Componentwise hierarchical penalties 
corresponding to the group lasso, group SCAD, and group MCP regularization have the respective forms
\begin{align} \label{Eq:componentwise_penalty}
    \mathcal{P}^{\text{GL}}_{\lambda,\text{C}}(\mathbf{A}) & = \lambda \sum_{i=1}^{d} \sum_{\ell = 1}^{p} \|\mathbf{A}_i^{(\ell:p)} \|_2, \quad \Omega_{i}(\mathbf{A}_{i}^{(\ell:p)}) = \lambda_{i} \|\mathbf{A}_i^{(\ell:p)} \|_2, \notag \\
    \mathcal{P}^{\text{AGL}}_{\lambda,\text{C}}(\mathbf{A})
    &=
    \lambda \sum_{i=1}^{d}\sum_{\ell=1}^{p}
    w_{i,\ell}^{C}
    \|\mathbf{A}_i^{(\ell:p)}\|_2, \quad
    \Omega_i(\mathbf{A}_i^{(\ell:p)})
    =
    \lambda_i w_{i,\ell}^{C}
    \|\mathbf{A}_i^{(\ell:p)}\|_2,
    \notag\\
    \mathcal{P}^{\text{SCAD}}_{\lambda,\text{C}}(\mathbf{A}) & = \sum_{i=1}^{d} \sum_{\ell = 1}^{p} P^{\text{SCAD}}_{\lambda,3.7}(\|\mathbf{A}_i^{(\ell:p)} \|_2), \quad \Omega_{i}(\mathbf{A}_{i}^{(\ell:p)}) = P^{\text{SCAD}}_{\lambda_{i},3.7}(\|\mathbf{A}_i^{(\ell:p)} \|_2),\notag \\
    \mathcal{P}^{\text{MCP}}_{\lambda,\text{C}}(\mathbf{A}) & = \sum_{i=1}^{d} \sum_{\ell = 1}^{p} P^{\text{MCP}}_{\lambda,3}(\|\mathbf{A}_i^{(\ell:p)} \|_2), \quad \Omega_{i}(\mathbf{A}_{i}^{(\ell:p)}) = P^{\text{MCP}}_{\lambda_{i},3}(\|\mathbf{A}_i^{(\ell:p)} \|_2),
\end{align}
where $\mathbf{A}_i^{(\ell:p)}$ is a $d(p-\ell + 1)$ dimensional row vector. 
Let $\mathbf{A}_{i,-i}^{(\ell)}$ be the $i^{th}$ row of the $d \times d$ matrix $\mathbf{A}^{(\ell)}$ without the $i^{th}$ element i.e., $\mathbf{A}_{i,-i}^{(\ell)} = \{\mathbf{A}_{ij}^{(\ell)}: j\neq i\}$ and let $\mathbf{A}_{i}^{([p+1]:p)} = \boldsymbol{0}$. 

The Own-other hierarchical penalties with respect to the group lasso, group SCAD and group MCP regularization are respectively
\begin{align} \label{Eq:own_other_penalty}
    \mathcal{P}^{\text{GL}}_{\lambda,\text{O}}(\mathbf{A}) & = \lambda \sum_{i=1}^{d} \sum_{\ell =1}^{p} \bigg[\|\mathbf{A}_i^{(\ell:p)}\|_2 +   \|\big(\mathbf{A}_{i,-i}^{(\ell)},\mathbf{A}_{i}^{([\ell + 1]:p)}\big)\|_2\bigg],  \notag \\
    \Omega_{i}(\mathbf{A}_{i}^{(\ell:p)}) & 
    = \lambda_{i} \bigg[\|\mathbf{A}_i^{(\ell:p)}\|_2 +   \|\big(\mathbf{A}_{i,-i}^{(\ell)},\mathbf{A}_{i}^{([\ell + 1]:p)}\big)\|_2\bigg], \notag \\
    \mathcal{P}^{\text{AGL}}_{\lambda,\text{O}}(\mathbf{A})
    &=
    \lambda\sum_{i=1}^{d}
    \Bigg[
    \sum_{\ell=1}^{p}
    w_{i,\ell}^{C}
    \|\mathbf{A}_i^{(\ell:p)}\|_2
    +
    \sum_{\ell=1}^{p-1}
    w_{i,\ell}^{O}
    \left\|
    \left(
    \mathbf{A}_{i,-i}^{(\ell)},
    \mathbf{A}_i^{([\ell+1]:p)}
    \right)
    \right\|_2
    \Bigg],
    \notag\\
    \Omega_i(\mathbf{A}_i^{(\ell:p)})
    &=
    \lambda_i
    \Bigg[
    w_{i,\ell}^{C}
    \|\mathbf{A}_i^{(\ell:p)}\|_2
    +
    \mathbb{I}_{\{\ell<p\}}w_{i,\ell}^{O}
    \left\|
    \left(
    \mathbf{A}_{i,-i}^{(\ell)},
    \mathbf{A}_i^{([\ell+1]:p)}
    \right)
    \right\|_2
    \Bigg],
    \notag\\
    \mathcal{P}^{\text{SCAD}}_{\lambda,\text{O}}(\mathbf{A}) & = \sum_{i=1}^{d} \sum_{\ell =1}^{p} \bigg[\mathcal{P}_{\lambda,3.7}^{\text{SCAD}}(\|\mathbf{A}_i^{(\ell:p)}\|_2) +   \mathcal{P}_{\lambda,3.7}^{\text{SCAD}}\Big(\|\big(\mathbf{A}_{i,-i}^{(\ell)},\mathbf{A}_{i}^{([\ell + 1]:p)}\big)\|_2\Big)\bigg],  \notag \\
    \Omega_{i}(\mathbf{A}_{i}^{(\ell:p)}) &= \bigg[\mathcal{P}_{\lambda_{i},3.7}^{\text{SCAD}}(\|\mathbf{A}_i^{(\ell:p)}\|_2) +   \mathcal{P}_{\lambda_{i},3.7}^{\text{SCAD}}\Big(\|\big(\mathbf{A}_{i,-i}^{(\ell)},\mathbf{A}_{i}^{([\ell + 1]:p)}\big)\|_2\Big)\bigg],  \notag\\
    \mathcal{P}^{\text{MCP}}_{\lambda,\text{O}}(\mathbf{A}) & = \sum_{i=1}^{d} \sum_{\ell =1}^{p} \bigg[\mathcal{P}_{\lambda,3}^{\text{MCP}}(\|\mathbf{A}_i^{(\ell:p)}\|_2) +   \mathcal{P}_{\lambda,3}^{\text{MCP}}\Big(\|\big(\mathbf{A}_{i,-i}^{(\ell)},\mathbf{A}_{i}^{([\ell + 1]:p)}\big)\|_2\Big)\bigg],  \notag \\
    \Omega_{i}(\mathbf{A}_{i}^{(\ell:p)}) & = \bigg[\mathcal{P}_{\lambda_{i},3}^{\text{MCP}}(\|\mathbf{A}_i^{(\ell:p)}\|_2) +   \mathcal{P}_{\lambda_{i},3}^{\text{MCP}}\Big(\|\big(\mathbf{A}_{i,-i}^{(\ell)},\mathbf{A}_{i}^{([\ell + 1]:p)}\big)\|_2\Big)\bigg]
\end{align}

The Elementwise hierarchical penalties with respect to group lasso, group SCAD, and group MCP are respectively 
{\color{black}
\begin{align} \label{Eq:elementwise_penalty}
    \mathcal{P}_{\lambda,\text{E}}^{\text{GL}}(\mathbf{A}) & = \lambda \sum_{i=1}^{d} \sum_{j=1}^{d} \sum_{\ell=1}^{p} \|\mathbf{A}_{ij}^{(\ell:p)} \|_2, \quad \Omega_{i}(\mathbf{A}_{i}^{(\ell:p)}) = \sum_{j=1}^{d}\lambda_{i}\|\mathbf{A}_{ij}^{(\ell:p)} \|_2, \notag \\
    \mathcal{P}^{\text{AGL}}_{\lambda,\text{E}}(\mathbf{A})
    &=
    \lambda \sum_{i=1}^{d}\sum_{j=1}^{d}
    \sum_{\ell=1}^{p}
    w_{ij,\ell}^{E}
    \|\mathbf{A}_{ij}^{(\ell:p)}\|_2, \quad
    \Omega_i(\mathbf{A}_i^{(\ell:p)})
    =
    \sum_{j=1}^{d}
    \lambda_i w_{ij,\ell}^{E}
    \|\mathbf{A}_{ij}^{(\ell:p)}\|_2,
    \notag\\
    \mathcal{P}_{\lambda,\text{E}}^{\text{SCAD}}(\mathbf{A}) & = \sum_{i=1}^{d} \sum_{j=1}^{d} \sum_{\ell=1}^{p} \mathcal{P}_{\lambda,3.7}^{\text{SCAD}}\big(\|\mathbf{A}_{ij}^{(\ell:p)}\|_2\big),\quad \Omega_{i}(\mathbf{A}_{i}^{(\ell:p)}) = \sum_{j=1}^{d}\mathcal{P}_{\lambda_{i},3.7}^{\text{SCAD}}\big(\|\mathbf{A}_{ij}^{(\ell:p)}\|_2\big), \notag \\
    \mathcal{P}_{\lambda,\text{E}}^{\text{MCP}}(\mathbf{A}) & = \sum_{i=1}^{d} \sum_{j=1}^{d} \sum_{\ell=1}^{p} \mathcal{P}_{\lambda,3}^{\text{MCP}}\big(\|\mathbf{A}_{ij}^{(\ell:p)}\|_2\big), \quad \Omega_{i}(\mathbf{A}_{i}^{(\ell:p)}) = \sum_{j=1}^{d}\mathcal{P}_{\lambda_{i},3}^{\text{MCP}}\big(\|\mathbf{A}_{ij}^{(\ell:p)}\|_2\big)
\end{align}}
Section~\ref{sec:penestalg} describes the 
steps to compute the 
penalized maximum likelihood estimates (MLEs) under each of the above scenarios.

\subsection{Implementation of penalized estimation}\label{sec:penestalg}






To compute the 
penalized MLE of $\boldsymbol{\Theta}$, we solve the objective function defined as 
\begin{equation}
    Q_{\lambda}(\boldsymbol{\Theta}) =  \Bigg\{-\frac{1}{T}\ell(\boldsymbol{A},\beta;\boldsymbol{X})+\sum_{i=1}^{d} \sum_{\ell=1}^{p}\Omega_{i}(\boldsymbol{A}_i^{(\ell:p)})\Bigg\}
\label{eq:objective_function}
\end{equation}
recall that $\boldsymbol{\Theta} = (\beta,\text{vec'}(\boldsymbol{A^{(1)}}),\ldots,\text{vec'}(\boldsymbol{A^{(p)}})$, and
$\ell(\boldsymbol{\Theta}; \boldsymbol{X})$ was shown in \eqref{Eq:simplog-likelihood} and $\Omega_{i}$ is defined in \eqref{Eq:componentwise_penalty}, \eqref{Eq:own_other_penalty} and \eqref{Eq:elementwise_penalty} for Componentwise, Own-other and Elementwise sparse structure respectively. The penalized MLE of $\boldsymbol{\Theta}$ can be given as
\begin{equation}\label{optim_1}
    \widehat{\boldsymbol{\Theta}}_{\lambda} \ =\
    \underset{\boldsymbol{\Theta}}{\text{arg min}}\ \Big\{ Q_{\lambda}(\boldsymbol{\Theta}) 
    \Big\},
\end{equation}

{\color{black}
We employ the block coordinate descent algorithm \citep{tseng2001convergence,bertsekas1999nonlinear} described in Section~\ref{sec:maxloglike} for initializing the parameters using their unpenalized maximum likelihood estimates. For a fixed value of $\beta$, the rows of $\boldsymbol{A}$ are updated sequentially by minimizing the penalized objective using the derivative-free \texttt{bobyqa} algorithm. Since the penalty is imposed only on the autoregressive coefficients, the conditional update of $\beta$ depends on $\boldsymbol{A}$. The penalized estimation procedure retains the alternating optimization method described in Section~\ref{sec:maxloglike}, with the hierarchical lag penalties incorporated into the row wise updates of $\boldsymbol{A}$.
Following a heuristic similar to that of \citet{cattivelli2020adaptive}, the upper reference value of the regularization parameter is defined as
\begin{equation*}    
\lambda_{\max} = \frac{
|\ell(\widetilde A,\widetilde\beta;X)|
}{
Tpd\nu
}
\end{equation*}
where ($\widetilde{\boldsymbol{A}},\widetilde\beta$) denotes the unpenalized maximum likelihood estimates and ($\nu>0$) is a scaling parameter. The factor ($pd$) corresponds to the number of autoregressive coefficients in each row optimization problem. A logarithmically spaced grid of ($K$) candidate values is constructed between $\lambda_{\max}$ and $\varepsilon_\lambda\lambda_{\max}$, where $0<\varepsilon_\lambda<1$.

For each candidate $\lambda$, the penalized estimator is computed using the procedure described above. The optimal penalty parameter is selected by minimizing the Bayesian information criterion (BIC) over the candidate grid, and the final parameter estimates correspond to the selected value of $\lambda$.
}

{\color{black}
\noindent\textbf{Convergence considerations:}
\citet{Basu2015a} establish non-asymptotic estimation error bounds for regularized estimators of stable Gaussian time series, demonstrating that rates comparable to those obtained under independent observations can be achieved, subject to factors reflecting temporal dependence. \citet{Basu2015b} further establish norm and variable-selection consistency for group-regularized vector autoregressive models. These results provide a theoretical motivation for hierarchical lag regularization, as introduced by \citet{nicholson2020high}, in which nested coefficient groups impose structured sparsity across autoregressive lags. Existing results for regularized high-dimensional time series motivate hierarchical penalization. Establishing estimation and selection consistency for the log-vMEM with multivariate Gamma errors, and convergence of the inexact block coordinate scheme, is beyond the scope of this paper and we leave this for future work.
}


\subsection{Simulated Data Analysis under HLag Penalties} \label{section:simulated_data_analysis}
\begin{figure}[H]
    \includegraphics[scale=0.45]{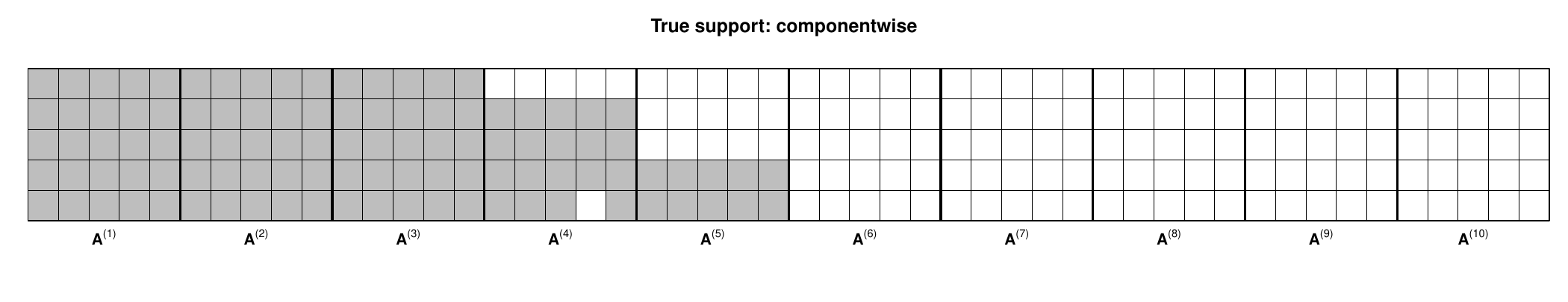} \\
      \includegraphics[scale=0.45]{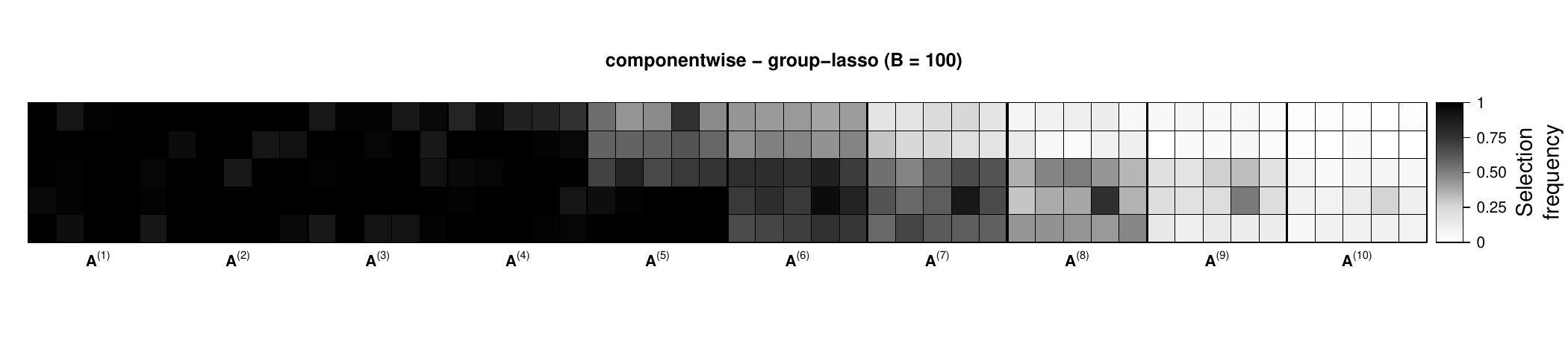} \\
      \includegraphics[scale=0.45]{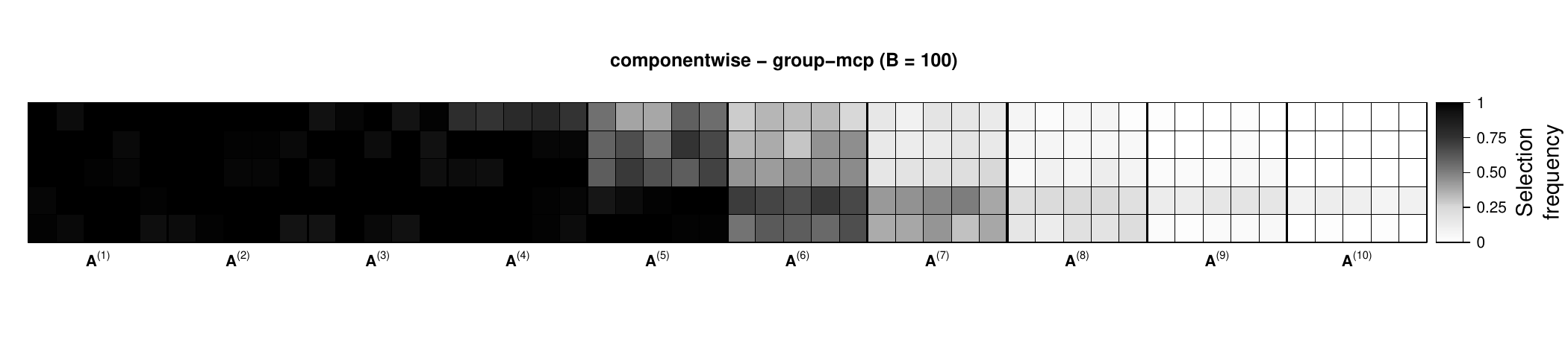} \\
      \includegraphics[scale=0.45]{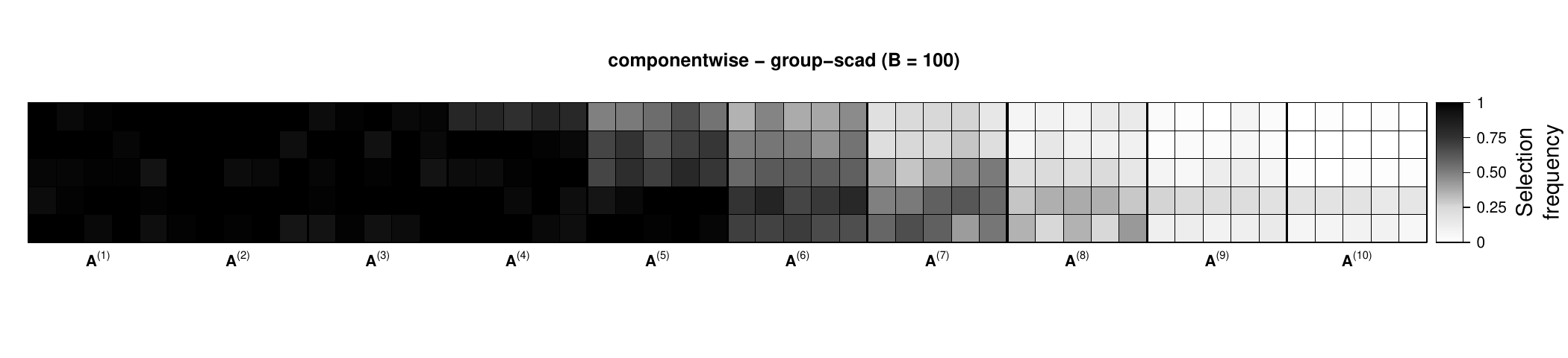}
    \caption{True active structures (row 1), simulated heatmaps under $B=100$ replicates for the estimated active structures under group lasso, group MCP, and group SCAD (rows 2-4) for the componentwise ($\mathbf{L}^{C}$) Hlag structure. \textcolor{black}{The sparsity structures show remarkable recovery after $B=100$ replications.}} 
    \label{fig:simul_activesparse_Lc}
 \end{figure}

\begin{figure}[H]
     \includegraphics[scale=0.45]{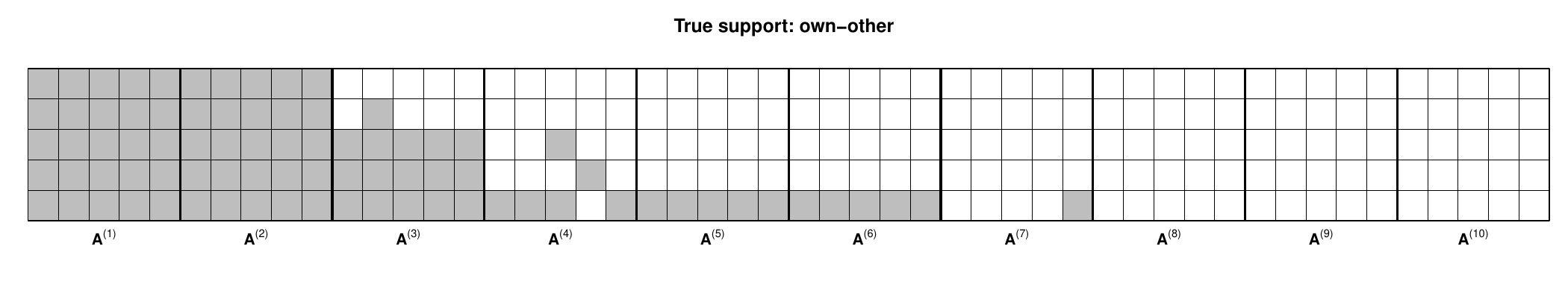} \\
      \includegraphics[scale=0.45]{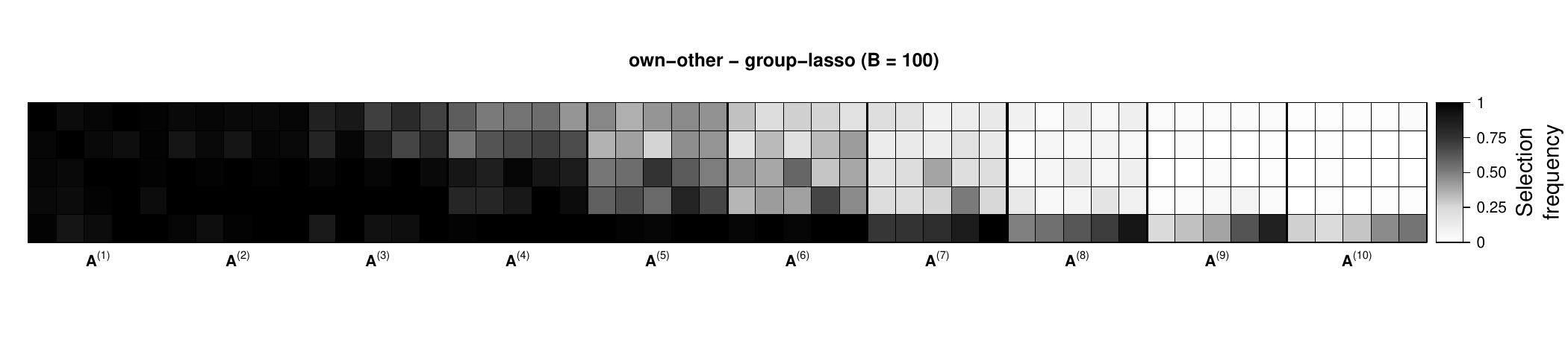} \\
      \includegraphics[scale=0.45]{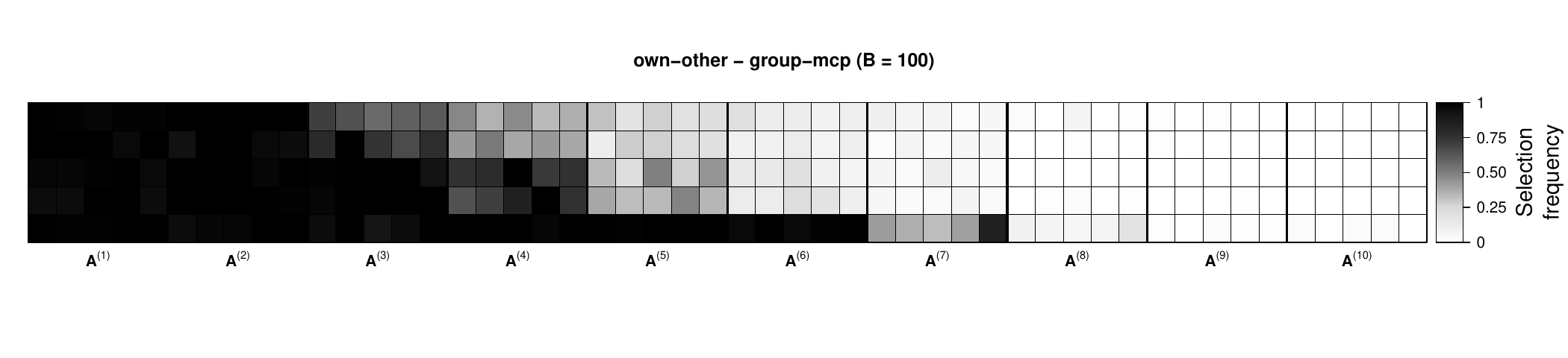} \\
     \includegraphics[scale=0.45]{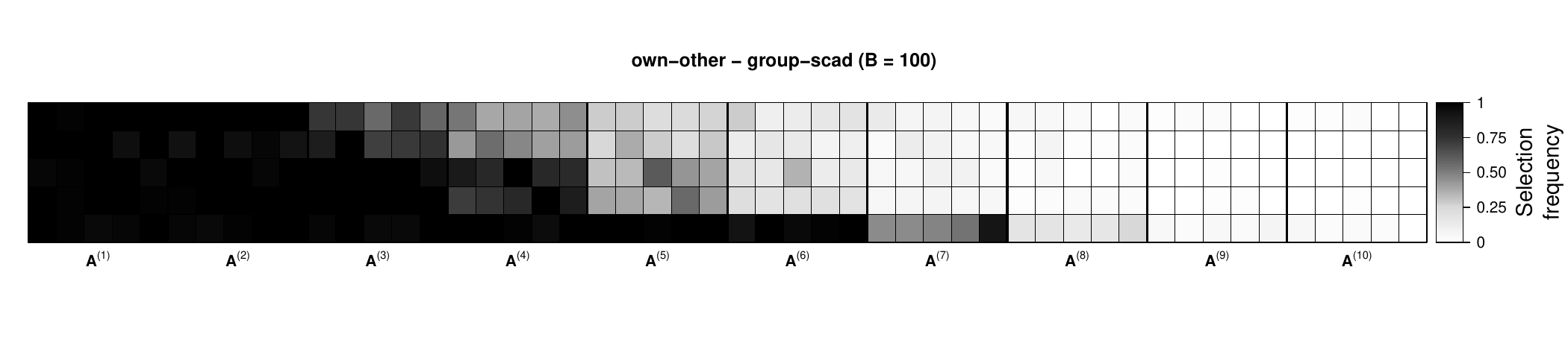}
    \caption{True active structures (row 1), simulated heatmaps under $B=100$ replicates for the estimated active structures under group lasso, group MCP, and group SCAD (rows 2-4) for the own-other ($\mathbf{L}^{O}$) Hlag structure. \textcolor{black}{The sparsity structures show remarkable recovery after $B=100$ replications.}} 
    \label{fig:simul_activesparse_Lo}
 \end{figure}

\begin{figure}[H]
     \includegraphics[scale=0.45]{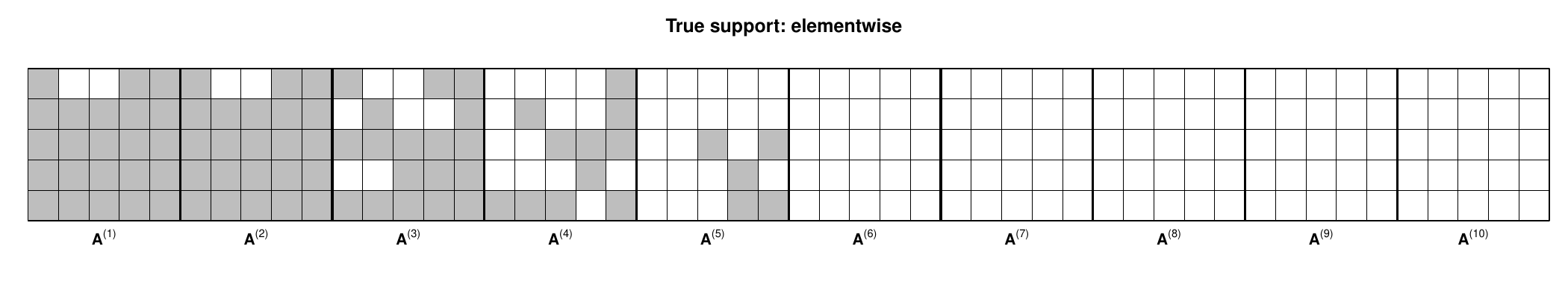}\\
      \includegraphics[scale=0.45]{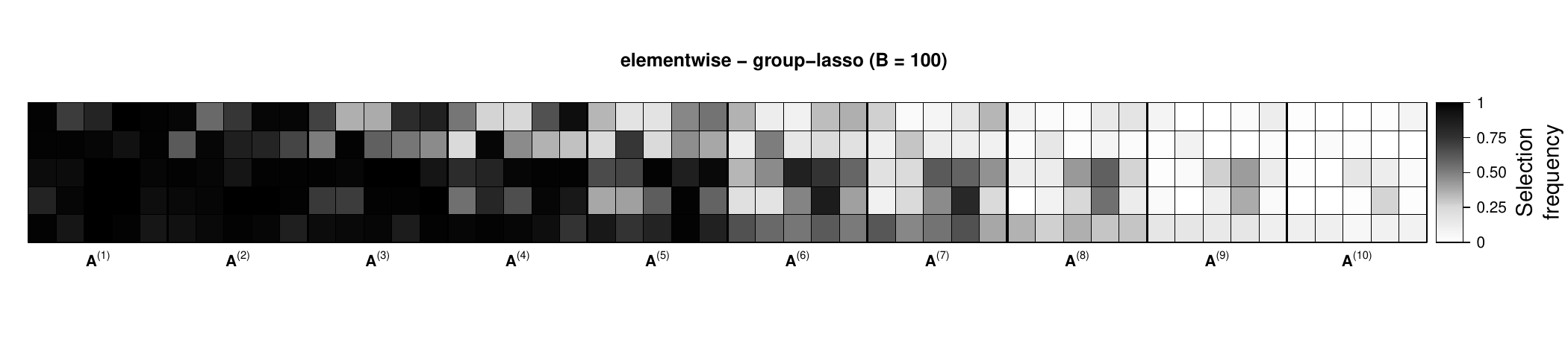} \\
        \includegraphics[scale=0.45]{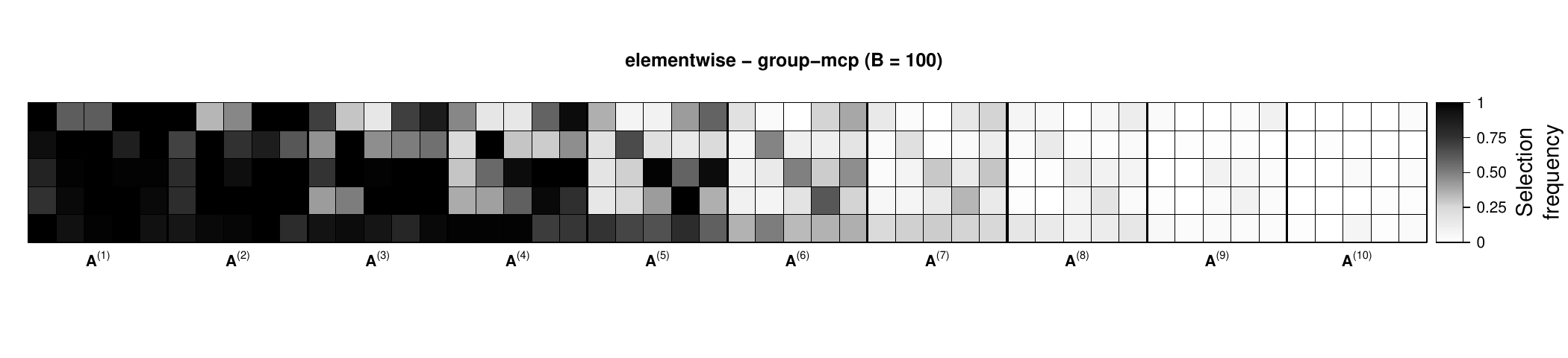} \\
        \includegraphics[scale=0.45]{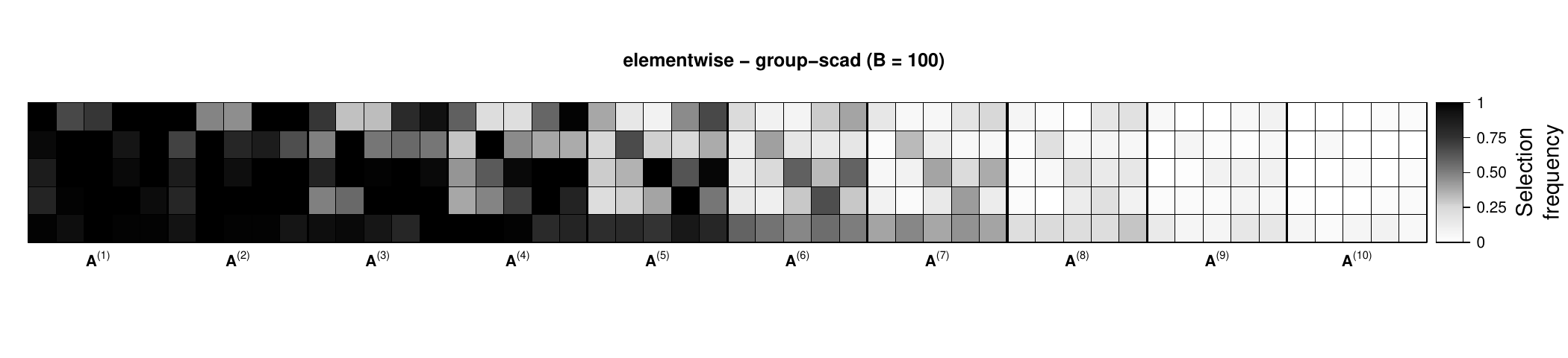}
        \caption{True active structures (row 1), simulated heatmaps under $B=100$ replicates for the estimated active structures under group lasso, group MCP, and group SCAD (rows 2-4) for the elementwise ($\mathbf{L}^{E}$) Hlag structure. \textcolor{black}{The sparsity structures show remarkable recovery after $B=100$ replications.}} 
        \label{fig:simul_activesparse_Le}
     \end{figure}

{\color{black}
We conduct a simulation study to assess  ability of the proposed penalized estimation procedure to recover the underlying autoregressive sparsity structure of the log-vMEM with multivariate Gamma errors. 
We consider the three hierarchical lag structures introduced in Section 3.1: componentwise, own-other, and elementwise. For each structure, we generate $B=100$ independent datasets of length $T=1500$, with dimension $d=5$ and maximum lag order $p=10$.

The true autoregressive coefficient matrices are constructed according to prespecified maximal-lag matrices. For each pair of components $(i,j)$, the entry $\mathbf{L}_{ij}$ specifies the maximum nonzero lag in the corresponding autoregressive coefficient sequence. The coefficient matrices are constructed to satisfy the stability condition of the log-vMEM process, following the companion-matrix approach discussed in \citet{lutkepohl2005new, wilms2021sparse}.
The maximal-lag matrices used for the three simulation scenarios are given in~\eqref{sim_lmatrices}.
\begin{equation}
    \mathbf{L}^{C}= \begin{pmatrix}
        3 & 3 & 3 & 3 & 3\\
        4 & 4 & 4 & 4 & 4\\
        4 & 4 & 4 & 4 & 4\\
        5 & 5 & 5 & 5 & 5\\
        5 & 5 & 5 & 5 & 5\\
\end{pmatrix}, \ 
    \mathbf{L}^{O}= \begin{pmatrix}
        2 & 2 & 2 & 2 & 2\\
        2 & 3 & 2 & 2 & 2\\
        3 & 3 & 4 & 3 & 3\\
        3 & 3 & 3 & 4 & 3\\
        6 & 6 & 6 & 6 & 7\\
\end{pmatrix}, \
  \mathbf{L}^{E}= \begin{pmatrix}
        3 & 0 & 0 & 3 & 4\\
        2 & 4 & 2 & 2 & 4\\
        3 & 3 & 5 & 4 & 5\\
        2 & 2 & 3 & 5 & 3\\
        4 & 4 & 4 & 5 & 5\\
\end{pmatrix}
\label{sim_lmatrices}
\end{equation}
For each simulated dataset, we estimate the penalized log-vMEM under the 
hierarchical lag structures using the group lasso, group MCP, and group SCAD penalties. The estimation follows the procedure described in Section~\ref{sec:penestalg}.
Figures~\ref{fig:simul_activesparse_Lc},~\ref{fig:simul_activesparse_Lo} and~ \ref{fig:simul_activesparse_Le} present the results for the componentwise, own-other, and elementwise structures, respectively. In each figure, the first row displays the true active coefficient structure, while the subsequent three rows display the heatmap of the empirical selection frequencies across the 100 simulation replications for the three penalties.

The results demonstrate reliable recovery of the HLag sparsity structures across all three simulation scenarios. 
Overall, 
they provide encouraging evidence of the ability of the proposed estimation procedure to recover structured autoregressive sparsity under different hierarchical lag specifications and a multivariate gamma likelihood.
}

\section{Analysis of Realized Volatility Measures}\label{emp_anal}
Several measures have been proposed in the literature for 
analyzing the volatility of an asset using intraday returns, including the daily range  \citep{parkinson1980extreme}, the realized volatility \citep{andersen2006volatility}, and the more recent realized kernels \citep{barndorff2009realized}, etc.
\cite{liu2015does} provides a good review of many realized estimators used to estimate realized variance of a financial asset. 

We analyze transaction level data for Microsoft (\texttt{NASDAQ: MSFT}) for the time period from January 4, 2010 to \textcolor{black}{December 31, 2015, consisting of $T= 1476$} trading days. We split each time series into a training portion consisting of  $T_1=1445$ observations (January 4, 2010 to November 5, 2015), and a test portion consisting of $T_2=31$ observations (November 6, 2015 to December 31, 2015).
On each day, the data consists of prices from each transaction. We sample the prices at 
sampling frequency $f$ by 
%
%
partitioning the trading time interval within a day (i.e., 9:30am to 4:00pm) into 
$S^{(f)}=390/f$
equal intervals, and recording the price  
$P^{(f)}_{s,t},~s=1,\ldots,S^{(f)}$ at the equidistant time points. For instance, 
when $f=5$, 
the sampling procedure records $S^{(f)}=78$  
MSFT prices, at 9:30am ($s=1$), at 9:35am ($s=2$), etc.
When $f=10$, we have $S^{(f)}=39$ prices, at
9:30 am ($s=1$), at 9:40 am ($s=2)$, etc. 
For prices sampled at frequency $f$, we compute the corresponding returns for MSFT on each day $t=1,\ldots,T$ as
\begin{align}    
r_{s,t}^{(f)} = \log(P^{(f)}_{s,t}) - \log(P^{(f)}_{s-1,t}),~s=1,\ldots,S^{(f)}
\end{align}

Let $\gamma^{(f)}_t(h),~h=1,\ldots,H$ denote the lag $h$ sample autocovariances of the returns $r^{(f)}_{s,t}$ constructed with sampling frequency $f$ on day $t$.

Based on the sampled returns, we construct 
multiple positive-valued time series which are popular daily robust measures of volatility. 
%
%
The {Bipower Variation} \citep{Barndorff2004}
isolates the variation of the pure price process and is 
\begin{equation*}
    \text{BPV}_t^{(f)} = \sum_{s=2}^{S^{(f)}} |r^{(f)}_{s,t}| |r^{(f)}_{s-1,t}|.
\end{equation*}
The \textit{median realized variance} on day $t$ 
is \citep{andersen2012jump} 
\begin{equation*}
    \text{MedRV}_t^{(f)} = \dfrac{\pi S^{(f)}}{(S^{(f)}-2)(6-4\sqrt{3}+\pi)} 
    \sum_{s=2}^{S^{(f)}-1} \text{median}\big(|r^{(f)}_{s-1,t}|,|r^{(f)}_{s,t}|,|r^{(f)}_{s+1,t}|\big)^2.
\end{equation*}

The \textit{realized kernels} estimate \citep{barndorff2009realized} on day $t$ with the 
modified Tukey-Hanning kernel function 
$k(y)= [1-\text{sin}(\pi/2 - \pi(1-y)^2)]/2$ is 
\begin{align*}
    \text{RK}^{(f)}_t & = \sum_{h= -H}^{H} k\Big(\frac{h}{H+1}\Big) \gamma^{(f)}_{t}(h). 
\end{align*}
Let $\overline{\text{BPV}}^{(f)}$,
$\overline{\text{MedRV}}^{(f)}$, and 
$\overline{\text{RK}}^{(f)}$
be the sample means of 
these measures across the $T_1$ days. 
We 
construct $d=5$ series $X_{ti},~j=1,\ldots,d$, defined by
\begin{align}
X_{t1} & = \frac{\sqrt{\text{MedRV}^{(5)}_t}}{\overline{\text{MedRV}}^{(5)}},
X_{t2} = \frac{\sqrt{\text{RK}^{(5)}_t}}{\overline{\text{RK}}^{(5)}}, 
X_{t3}  = \frac{\sqrt{\text{BPV}^{(5)}_t}}{\overline{\text{BPV}}^{(5)}},
X_{t4} = \frac{\sqrt{\text{MedRV}^{(10)}_t}}{\overline{\text{MedRV}}^{(10)}}, 
X_{t5} = \frac{\sqrt{\text{RK}^{(10)}_t}}{\overline{\text{RK}}^{(10)}}
\label{eq:Xt_eqs}
\end{align}
 
 
\begin{figure}[H]
    \centering
\includegraphics[scale=0.15]{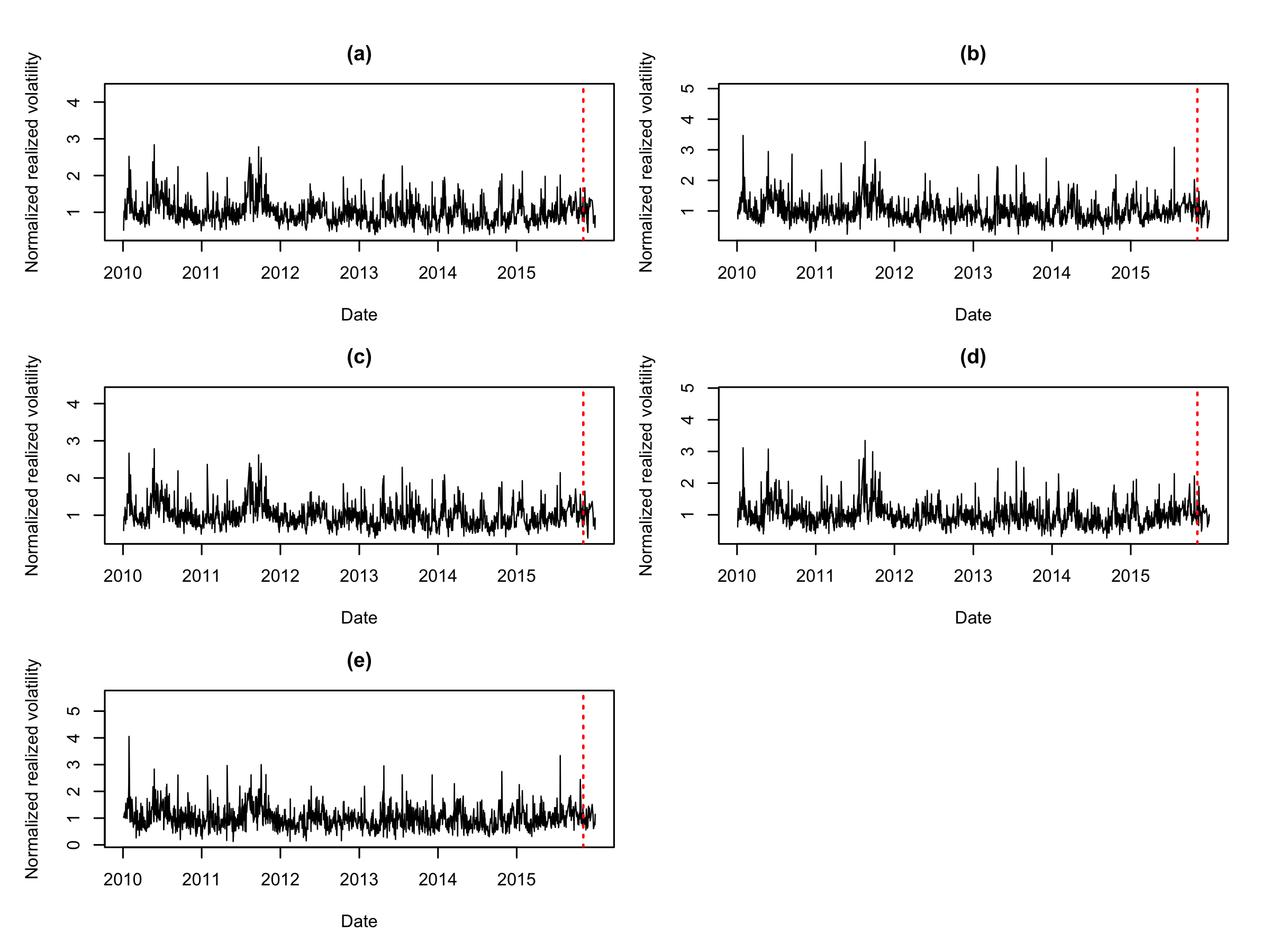}
    \caption{Realized volatility measures for \texttt{NASDAQ:MSFT} from January 4, 2010 to November 5, 2015. Panels (a)--(e) show the normalized 5-minute median realized volatility (\(X_{t1}\)), 5-minute realized kernel (\(X_{t2}\)), 5-minute bipower variation (\(X_{t3}\)), 10-minute median realized volatility (\(X_{t4}\)), and 10-minute realized kernel (\(X_{t5}\)) respectively.}
\label{fig:realized_measures_MSFT}
\end{figure}

%
\textcolor{black}{
Figure \ref{fig:realized_measures_MSFT} shows plots of the transformed realized volatility measures
$X_{tj},~j=1,\ldots,5$, while Table \ref{Table:Descriptive Statistics} shows their respective 
standard deviations, skewness, kurtosis and lag 1 autocorrelations. All the series are temporally dependent, positively skewed and heavy-tailed.
The five realized volatility measures exhibit moderate to high positive Kendall correlations, ranging from 0.52 between $X_{t1}$ and $X_{t5}$ to 0.88 between $X_{t1}$ and $X_{t3}$.}  

\textcolor{black}{We split each time series into a training portion consisting of  $T_1=1445$ observations (January 4, 2010 to November 5, 2015), and a test portion consisting of $T_2=31$ observations (November 6, 2015 to December 31, 2015).}
For analyzing the real data, 
we fit the model in \eqref{Eq:simplog-likelihood}  with $p = 10$ to the training data, under three different hierarchical lag structures (i.e.,  componentwise, own-other and elementwise),
and four different penalties (i.e., group lasso, adaptive group-lasso, group MCP, and group SCAD). \textcolor{black}{
Therefore, these competing models are trained on the data from January 4, 2010 to November 5, 2015, and forecasted on the first five observations in the test data from November 6, 2015 to November 12, 2015. Note that November 7 and November 8 are not trading days.
We fit each of these twelve combinations on eight different $\lambda$ choices, and the best is chosen by smallest insample BIC.
This task is parallelized over 96 cores with 350GB of memory, and the recorded runtime is 17.8 hours for a single origin (five trading days forecast window).
The model having the least forecast error among the competing models has the Own-Other HLag structure and Adaptive Group Lasso penalty.}
\begin{table}[H]
\centering
\begin{tabular}{|c|c|c|c|c|c|}
\hline
 Measure & St.Dev. & Skewness & Kurtosis & AutoCorr(1) \\ \hline
 $X_{t1}$ & 0.33 & 1.36 & 6.04 & 0.48 \\ 
$X_{t2}$ & 0.38 & 1.55 & 7.86 & 0.38 \\ 
$X_{t3}$ & 0.33 & 1.35 & 5.94 & 0.50 \\ 
$X_{t4}$ & 0.37 & 1.52 & 7.37 & 0.41 \\ 
$X_{t5}$ & 0.42 & 1.40 & 7.47 & 0.29 \\ \hline
\end{tabular}
\caption{Descriptive statistics for $X_{tj},~j=1,\ldots,5$. The five volatility measures demonstrate positive skew and high kurtosis.}
\label{Table:Descriptive Statistics}
\end{table}

\begin{figure}[H]
    \centering
\includegraphics[scale=0.15]{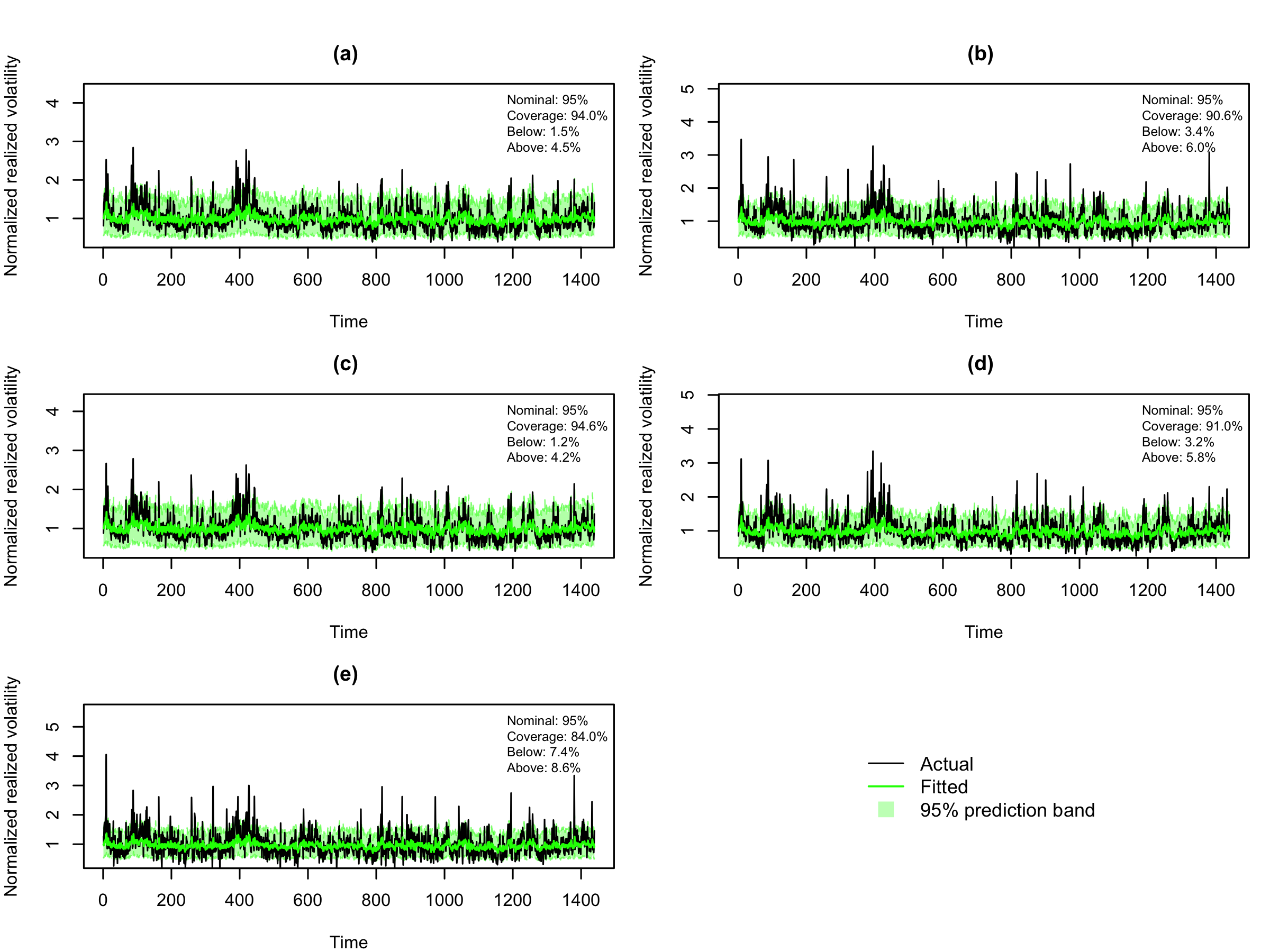}
    \caption{Plots of the actual vs fitted values for the insample data. (a)-(e) shows the plots for $X_{tj}$ (in black) and $\hat{X}_{tj}=\hat{\mu}_{tj}$ (in green),$~j=1,\ldots,5$. The 95\% prediction bands denote adequate coverage except for some outliers.}
\label{fig:fitted_plots_MSFT}
\end{figure}

\begin{figure}[H]
    \centering
\includegraphics[scale=0.15]{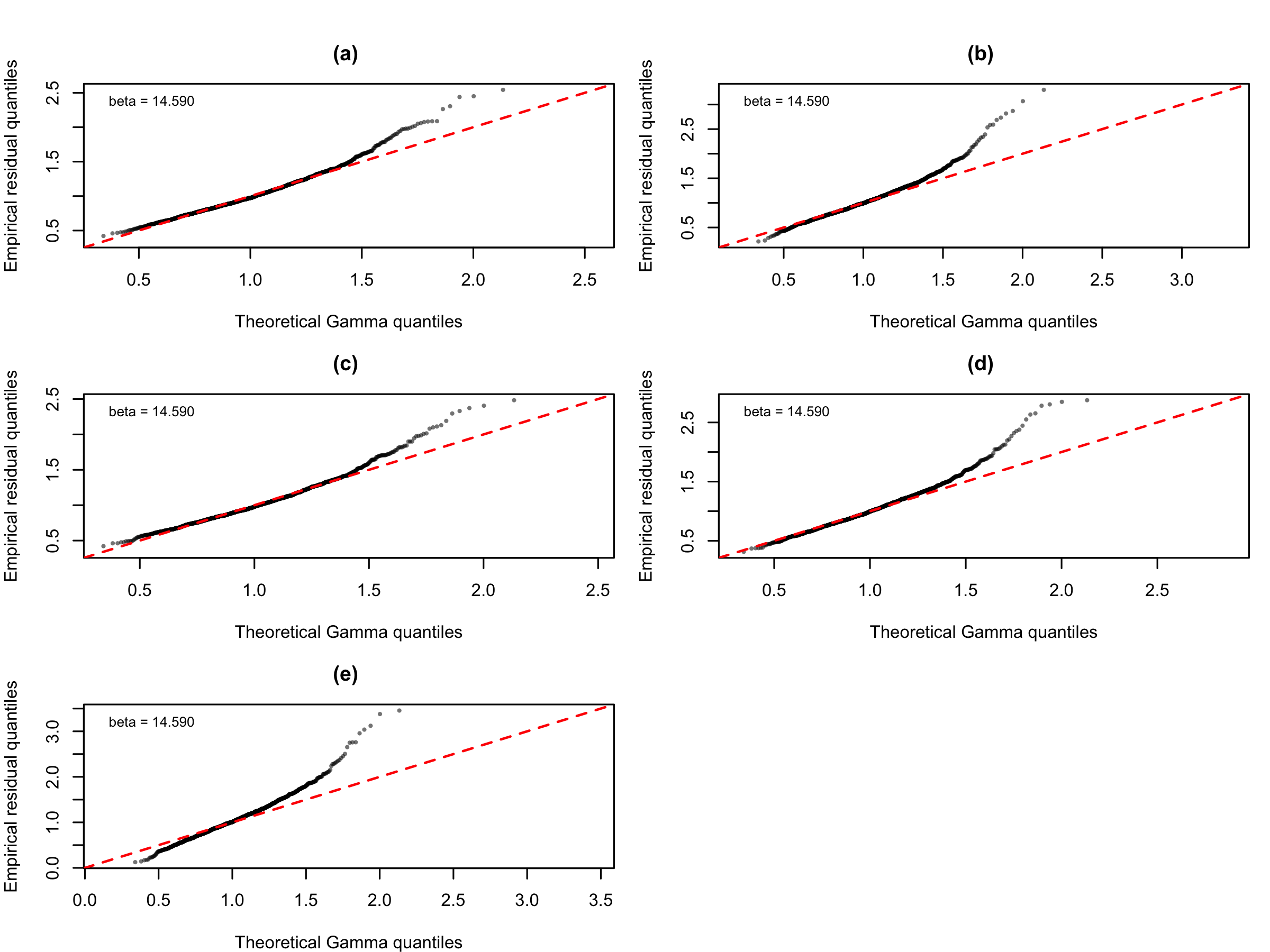}
    \caption{Gamma QQ plots for the multiplicative residuals of the five realized volatility measures. Panels (a)--(e) compare the empirical quantiles of the  insample multiplicative residuals, $\widehat{\epsilon}_{tj}= X_{tj}/\widehat{\mu}_{tj}$, $j=1,\ldots,5$, with the corresponding theoretical Gamma quantiles under the fitted values. The dashed red line represents the reference line for agreement between the empirical and theoretical quantiles. The residuals exhibit reasonable agreement with the assumed Gamma distribution over the central quantiles, with departures becoming more pronounced in the upper tail.}
\label{fig:residual_qqplot_uncensored}
\end{figure}

\begin{figure}[H]
    \centering
\includegraphics[scale=0.15]{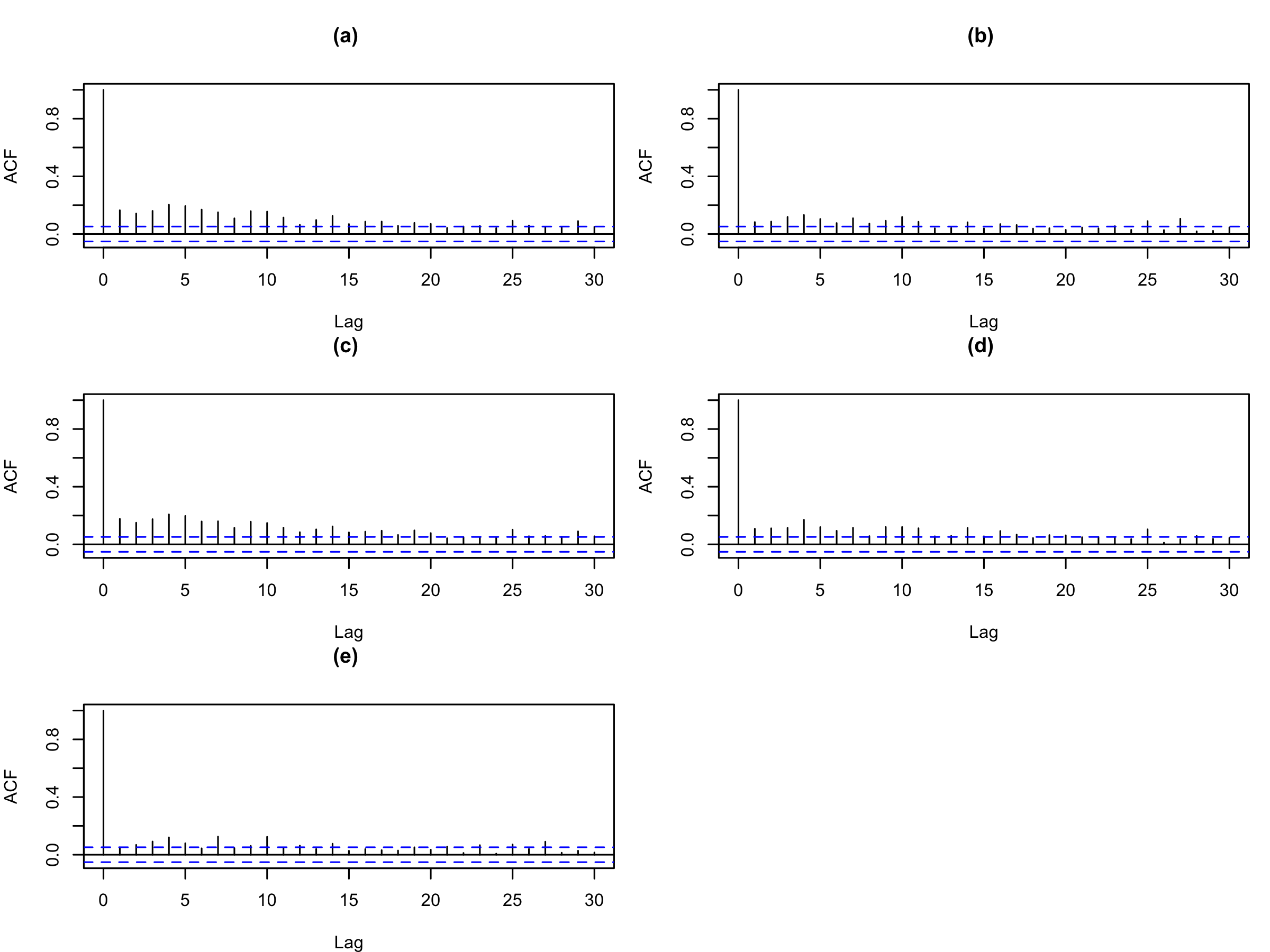}
    \caption{Autocorrelation functions of the multiplicative residuals for the five realized volatility measures. Panels (a)--(e) show the insample ACFs of the multiplicative residuals, $\widehat{\epsilon}_{tj}=X_{tj}/\widehat{\mu}_{tj}$, $j=1,\ldots,5$, up to lag 30. The blue dashed lines represent the approximate 95\% confidence bounds for zero autocorrelation. The residual series exhibit substantially reduced temporal dependence, although some statistically significant autocorrelation remains at several lags.}
\label{fig:residual_plots_MSFT_1}
\end{figure}

{\color{black}

As shown in Figure~\ref{fig:fitted_plots_MSFT}, the selected model provides an adequate fit to the five realized volatility series, with the fitted conditional means capturing the overall temporal dynamics of the observed series. The empirical coverage of the 95\% prediction bands ranges from 84.0\% to 94.6\%, with most departures associated with high-volatility peaks. The residual diagnostics further support the overall goodness of fit. The autocorrelation functions in Figure~\ref{fig:residual_plots_MSFT_1} show substantially reduced temporal dependence in the multiplicative residuals, although some residual autocorrelation remains. The Gamma QQ plots
in Figure~\ref{fig:residual_qqplot_uncensored} indicate a reasonable distributional fit over the central part of the distribution, with departures primarily in the upper tail.

From a computational perspective, the model-selection procedure which involves model fitting, optimal $\lambda$ selection, and forecasting, 
requires approximately 17.8 hours of clock time for one origin.
{In practice, model selection is rarely done based on a single set of forecasts at a fixed origin, we need a rolling window analysis with multiple origins. The current computational runtime for this task would take approximately 480.06 hours $\approx$ 20 days for generating 27 origin rolling forecasts.}
These heavy computational deterrents 
motivate us to identify the principal computational bottleneck and develop GPU-based
techniques to accelerate the model estimation.
These details are explored in the next section.
}

\section{Computational Acceleration Using GPU-Based Quadrature Integration}\label{sec:gpu}
A major computational bottleneck in estimating the parameters of the proposed log-vMEM arises from the numerical integration used in the multivariate 
Gamma log-likelihood. The final term in the log-likelihood 
function in \eqref{Eq:simplog-likelihood} requires, for each time point $t$, evaluation of the 
one-dimensional integral



\begin{align}
\label{eq:gpu_integral_1}
\mathcal{I}_t =
\int_{0}^{B_t}
\bigg\{\prod_{i=1}^{d}
\left(
\frac{x_{ti}}{\mu_{ti}}-z_t
\right)^{\big(\frac{\beta}{2}-1\big)}\bigg\}
\times
z_t^{\big(\frac{\beta}{2}  - 1 \big)}\times {\color{black}e^{\beta(d-1)z_t}}
\end{align}

\begin{table}[H]
\centering
\resizebox{\textwidth}{!}{%
\begin{tabular}{|l|ccccc|ccccc|ccccc|}
\hline\hline
&
\multicolumn{5}{c|}{\textbf{MSE}} &
\multicolumn{5}{c|}{\textbf{MAD}} &
\multicolumn{5}{c|}{\textbf{MAPE}} \\
\cline{2-16}

\textbf{Model}
& $h=1$ & $h=2$ & $h=3$ & $h=4$ & $h=5$
& $h=1$ & $h=2$ & $h=3$ & $h=4$ & $h=5$
& $h=1$ & $h=2$ & $h=3$ & $h=4$ & $h=5$ \\
\hline

No penalty
& 0.0911 & 0.104 & 0.107 & 0.112 & 0.109
& 0.234 & 0.259 & 0.262 & 0.270 & 0.263
& 0.273 & 0.318 & 0.324 & 0.337 & 0.333 \\

Componentwise-Adaptive Group Lasso
& 0.0819 & 0.0873 & 0.0886 & 0.0907 & 0.0848
& 0.215 & 0.233 & 0.233 & 0.240 & 0.231
& 0.246 & 0.281 & 0.284 & 0.294 & 0.285 \\

Componentwise-Group Lasso
& 0.0823 & 0.0882 & 0.0919 & 0.0942 & 0.0915
& 0.219 & 0.233 & 0.238 & 0.245 & 0.240
& 0.252 & 0.282 & 0.291 & 0.302 & 0.299 \\

Componentwise-Group MCP
& 0.0892 & 0.0992 & 0.105 & 0.105 & 0.107
& 0.229 & 0.251 & 0.259 & 0.259 & 0.257
& 0.267 & 0.306 & 0.320 & 0.322 & 0.325 \\

Componentwise-Group SCAD
& 0.0895 & 0.104 & 0.101 & 0.111 & 0.108
& 0.233 & 0.257 & 0.255 & 0.270 & 0.260
& 0.268 & 0.315 & 0.314 & 0.336 & 0.330 \\

Elementwise-Adaptive Group Lasso
& 0.0857 & 0.0835 & 0.0830 & 0.0901 & 0.0890
& 0.225 & 0.225 & 0.225 & 0.236 & 0.236
& 0.261 & 0.271 & 0.272 & 0.291 & 0.291 \\

Elementwise-Group Lasso
& 0.0821 & 0.0867 & 0.0899 & 0.0924 & 0.0894
& 0.218 & 0.231 & 0.233 & 0.241 & 0.237
& 0.250 & 0.279 & 0.285 & 0.297 & 0.294 \\

Elementwise-Group MCP
& 0.0892 & 0.103 & 0.106 & 0.108 & 0.104
& 0.235 & 0.255 & 0.258 & 0.265 & 0.257
& 0.272 & 0.312 & 0.319 & 0.328 & 0.324 \\

Elementwise-Group SCAD
& 0.0871 & 0.102 & 0.104 & 0.106 & 0.106
& 0.229 & 0.257 & 0.258 & 0.262 & 0.260
& 0.264 & 0.314 & 0.317 & 0.324 & 0.327 \\

\textbf{Own-Other-Adaptive Group Lasso}
& \textbf{0.0809} & \textbf{0.0832} & \textbf{0.0837} & \textbf{0.0869} & \textbf{0.0836}
& \textbf{0.215} & \textbf{0.226} & \textbf{0.226} & \textbf{0.233} & \textbf{0.229}
& \textbf{0.244} & \textbf{0.270} & \textbf{0.273} & \textbf{0.284} & \textbf{0.280} \\

Own-Other-Group Lasso
& 0.0810 & 0.0858 & 0.0893 & 0.0919 & 0.0889
& 0.216 & 0.230 & 0.234 & 0.241 & 0.237
& 0.248 & 0.277 & 0.285 & 0.297 & 0.293 \\

Own-Other-Group MCP
& 0.0889 & 0.0987 & 0.102 & 0.105 & 0.105
& 0.231 & 0.250 & 0.254 & 0.262 & 0.259
& 0.267 & 0.304 & 0.312 & 0.325 & 0.324 \\

Own-Other-Group SCAD
& 0.0897 & 0.0955 & 0.103 & 0.102 & 0.0981
& 0.234 & 0.248 & 0.255 & 0.257 & 0.251
& 0.269 & 0.300 & 0.316 & 0.318 & 0.313 \\

\hline\hline
\end{tabular}%
}
\caption{Out-of-sample forecast accuracy for the rolling-window GPU analysis.
For each rolling forecast origin, the tuning parameter is selected using the
Bayesian information criterion (BIC) based only on the corresponding training
window. Reported values are forecast accuracy measures averaged across rolling
origins and the five realized volatility measures at horizons
$h=1,\ldots,5$.}
\label{tab:rolling_gpu_forecast_accuracy}
\end{table}

This computation becomes
particularly expensive during parameter estimation because the BOBYQA
optimizer described in Section~\ref{sec:maxloglike} repeatedly evaluates the log-likelihood
at different trial parameter values. Thus, although each integral is only
one-dimensional, numerical integration is performed a large number of times
over the course of the block coordinate descent algorithm.

An alternative technique to the numerical integration can be the  fixed-node Gauss-type quadrature. Let $z_{tk}$ and $w_{tk}$, $k=1,\ldots,K$, denote
appropriately transformed quadrature nodes and weights on $[0,B_t]$. Then
the integral above can be approximated by
\begin{align}
\label{eq:gpu_integral_2}
\tilde{\mathcal{I}}_t
=
\sum_{k=1}^{K} w_{tk}
\bigg\{\prod_{i=1}^{d}
\left(
\frac{x_{ti}}{\mu_{ti}}-z_{tk}
\right)^{\big(\frac{\beta}{2}-1\big)}\bigg\}
\times
z_{tk}^{\big(\frac{\beta}{2}  - 1 \big)}\times {\color{black}e^{\beta(d-1)z_{tk}}}
\end{align}
The calculations for different
$t$ and quadrature nodes $k$ are independent conditional on the model
parameters. They can therefore be vectorized and evaluated simultaneously,
making the likelihood particularly suitable for massively parallel
computation on a graphics processing unit (GPU).

To examine how the computational advantage of GPU-based likelihood
evaluation changes with the dimension of the system, we conducted a
scaling experiment with $T=1500$ and $K=64$ quadrature nodes.
In the GPU
implementation, the data and quadrature quantities were transferred to
GPU memory once and retained there across repeated likelihood evaluations,
thereby avoiding repeated data-transfer overhead. Table~\ref{tab:gpu_scaling}
reports the computation time per likelihood evaluation on CPUs vs GPUs.

\begin{table}[H]
\centering
\begin{tabular}{|r|r|r|r|}
\hline
{$\boldsymbol{d}$} & \textbf{CPU (sec)} & \textbf{GPU (sec)} & \textbf{CPU/GPU} \\
\hline
5   & 0.050 & 0.022 & 2.27 \\
50  & 0.103 & 0.034 & 3.03 \\
100 & 0.195 & 0.047 & 4.15 \\
200 & 0.462 & 0.055 & 8.40 \\
500 & 1.095 & 0.064 & 17.11 \\
\hline
\end{tabular}
\caption{Scaling of CPU and GPU likelihood evaluation with dimension ($d$, $T=1500$, $K=64$, and $p=10$). The evaluation times are recorded in seconds, and the final column demonstrates the effective gain of the GPU over CPU. \textcolor{black}{The GPU computations for this table were performed using the integrated GPU of the Apple M4 Pro chip.
}}
\label{tab:gpu_scaling}
\end{table}

For low dimensional models, the computational gains are not significant. This is because the workload is insufficient
to offset the overhead associated with GPU execution. For example, at
$d=5$, GPU evaluation remains faster by a factor of $ 2.27$ only. The advantage of GPU computation,
however, increases rapidly with dimension. At $d=50$, the GPU
implementation is approximately $3$ times faster than the CPU
implementation, increasing to $4.15$ times at $d=100$, $8.4$ times at
$d=200$, and $17.1$ times at $d=500$. This scaling behavior is
extremely pertinent for applications involving large systems of
financial assets. For example, modeling a system on the scale of the
\texttt{S\&P 500} would involve approximately 500 component time series, a setting in
which repeated likelihood evaluation can benefit substantially from GPU
parallelization.
\textcolor{black}{Another application for the GPU-accelerated parallelization would be for a simulation run when $d=500$, $p=50$, and to test parameter recovery.}

{\color{black} To compare the competing models, which is one of the stated goals of this study, we use the GPU accelerated likelihood to perform a rolling window moving origin validation procedure on the real data. Table~\ref{tab:rolling_gpu_forecast_accuracy} demonstrates the rolling window performance metrics averaged over 27 origins. 
We evaluate the out-of-sample forecasting performance of the competing penalty and HLag
specifications using the $h$-step-ahead($h=1,2,\ldots 5$) mean squared error (MSE), mean absolute deviation (MAD),
and mean absolute percentage error (MAPE), averaged across the five realized volatility measures.
It is observed that Own-Other HLag structure with a Adaptive Group Lasso penalty performs the best in terms of out sample rolling window forecasting measures.
The computational runtime of this procedure is recorded as 158.79 hours $\approx$ 6.62 days. This corroborates the computational acceleration aspect of the proposed method, and provides a $3\times$ speedup as compared to the CPU method.

The simulation runs reported in this paper use
the CPU-based implementation to ensure that the proposed methodology remains
reproduceable on standard computing environments without
the need for any specialized hardware. 
The GPU version is primarily used for the rolling origin validation run here to ensure computational acceleration.}
We provide both the GPU and the CPU implementations of the
log-likelihood in the accompanying \href{https://github.com/Rohan-Chhatre/sparse_logvMEM}{GitHub repository}.

\section{Conclusion} \label{section:concl}
In this paper, we describe sparse log-vMEMs for modeling multivariate positive-valued time series. We introduce the multivariate Gamma error distribution which facilitates 
contemporaneous correlation among the error terms. 
The common-factor Gamma specification imposes a
constant pairwise correlation among the multiplicative errors. Allowing more flexible dependence structures,
while retaining computational tractability, is a
potential direction for future research.
A block co-ordinate descent algorithm \citep{wright2015coordinate} is used for optimizing the likelihood function to ensure quicker convergence relative to a joint maximizing of the full likelihood.
As the dimension of the time series grows, log-vMEMs are affected by the \textit{curse of dimensionality} issue. To mitigate this issue, we adopt HLag structures \citep{nicholson2020high} and expand its usage from Gaussian VARs to log-vMEMs, allowing for flexible lag selection for regularization. Further, we expand the modeling tool from convex group lasso to non-convex penalties such as group SCAD and group MCP.
\textcolor{black}{The model demonstrates successful recovery of the true parameters in the simulation results both in the penalized, and the unpenalized runs. The successful simulation runs provide adequate justification of the model performance.}

\textcolor{black}{The real data analysis conducted for the five volatility measures of \texttt{NASDAQ:MSFT} demonstrate a decent 
fit on the in-sample data
with an issue in modeling the of the peaks. After accessing the residuals of the model, we observe a heavy tail presence in all five volatility measures at varying extent.
Upon conducting an empirical experiment of truncating the data at $95^{\text{th}}$ quantile for each $X_{tj} \ ;j=1\ldots5$, it is observed that the model is a good fit to the data, and the heavy upper tail ceases to exist in the qqplot of the residuals.
This can be viewed as a limitation of this paper 
and a resolution to this problem can be presented proposing a mixture error distribution
in the log-vMEM model which can take care of both the heavy \textit{tail} and the \textit{bulk}.}

Another key contribution of this study is the 
GPU-acceleration. We observe a $3\times$ speed up for the model fitting procedure after adopting a GPU-accelerated numerical quadrature integration technique. 
This computational speed up enables us to evaluate the competing models holistically, by performing a rolling window validation technique, and averaging out the forecast errors. The Own-Other HLag structure with a Adaptive Group Lasso penalty performs the best in terms of the out sample forecasting measures.


Our results show the promise of GPU-accelerated regularized modeling for positive-valued time series. \textcolor{black}{We also provide a discussion about the computational acceleration using GPUs which can help scale such models. This discussion would be helpful in extending this model to spatio-temporal settings in 
answering environmental, meterological questions, etc.}
Another direction for future research could involve exploring Bayesian shrinkage priors instead of penalized regression, which can also help with quantifying uncertainty of the regression coefficients. \\





\appendix
{\color{black}
\section{Appendix}
\subsection{Normalization issue in the existing \citet{tsionas2004bayesian} formulation}
\label{app:normalization_issue}

Consider the random vector $\mathbf{X} = [X_1,X_2,\ldots X_d]$ having the following representation
\begin{equation}    
X_i=V_i+Z,\qquad i=1,\ldots,d,
\label{eq:tsionas_rv}
\end{equation}
where $V_1,\ldots,V_d$ are independent $\operatorname{Gamma(\lambda_i,\theta_i); \ (i=1,2\ldots d)}$, and $Z$ is an independent $\operatorname{Gamma(\alpha,\beta)}$. We examine the expression appearing in Equation~(4) of \citet{tsionas2004bayesian}, 
\begin{equation}
\begin{aligned}
f_T(\mathbf{x})
&=
\frac{\beta^\alpha}{\Gamma(\alpha)}
\prod_{i=1}^{d}
\frac{\theta_i^{\lambda_i}}{\Gamma(\lambda_i)}
\prod_{i=1}^{d}
\left\{
\int_0^{B}
(x_i-z)^{\lambda_i-1}
e^{-\theta_i(x_i-z)}
z^{\alpha-1}e^{-\beta z}\,dz
\right\},
\end{aligned}
\label{eq:appendix_tsionas_density}
\end{equation}
where $B=\min_{1\leq i\leq d}x_i$,
$\mathbf{x}\in(0,\infty)^d$, and all parameters are positive.
For \eqref{eq:appendix_tsionas_density}, to define a valid probability density, it must satisfy
\[
\int_{(0,\infty)^d}f_T(\mathbf{x})\,d\mathbf{x}=1.
\]

\begin{proposition}
\label{prop:normalization}
For every integer $d\geq 2$, the expression in
\eqref{eq:appendix_tsionas_density} does not define
a valid probability density for all admissible
parameter values. In particular, setting
\[
\alpha=\beta=\lambda_i=\theta_i=1,
\qquad i=1,\ldots,d,
\]
yields
\[
\int_{(0,\infty)^d}
f_T(\mathbf{x})\,d\mathbf{x}
=\frac{d!}{d^d}<1
\]
\end{proposition}

\begin{proof}
Under the given parameter configuration, all
multiplicative constants in
\eqref{eq:appendix_tsionas_density} equal one. Consequently, the p.d.f. $f_T$ is
\begin{align*}
f_T(\mathbf{x})
&=
\prod_{i=1}^{d}
\left\{
\int_0^B e^{-(x_i-z)}e^{-z}\,dz
\right\}\\
&=
\prod_{i=1}^{d}
\left\{
e^{-x_i}\int_0^B dz
\right\}\\
&=
B^d\exp\left(-\sum_{i=1}^{d}x_i\right)
\end{align*}
The resulting expression is symmetric in
$x_1,\ldots,x_d$, , since both
$\sum_{i=1}^d x_i$ and $B=\min_{1\leq i\leq d}x_i$ remain unchanged under any permutation of the coordinates. Consider the region in which $x_1=m=\min_i x_i$. On this region, we have $B=m, \ x_1=m, \ x_i \in (m,\infty); \ i = 2,3\ldots d$. 
Let,
\begin{align}
I_T(d)
&:=\int_{(0,\infty)^d}f_T(\mathbf{x})\,d\mathbf{x} \nonumber \\
&=d\int_0^\infty m^d e^{-m}
\prod_{i=2}^d
\left(\int_m^\infty e^{-x_i}\,dx_i\right)dm \nonumber \\ 
&=d\int_0^\infty m^d e^{-m}
\prod_{i=2}^d e^{-m}\,dm \nonumber \\ 
&=d\int_0^\infty m^d e^{-m}e^{-(d-1)m}\,dm \nonumber \\ 
&=d\int_0^\infty m^d e^{-dm}\,dm \nonumber \\ 
&=d\frac{\Gamma(d+1)}{d^{d+1}} \nonumber \\
&=\frac{d!}{d^d}
\end{align}
For every integer $d\geq2$,
\[
d!=\prod_{j=1}^{d}j
<
\prod_{j=1}^{d}d=d^d.
\]
It follows that
\[
I_T(d)=\frac{d!}{d^d}<1,
\]
contradicting the normalization requirement for a
probability density.
\end{proof}

This counterexample establishes that the expression in \eqref{eq:appendix_tsionas_density} does not, in general, define a normalized multivariate probability density.

The density propsed by \citet{tsionas2004bayesian} in \eqref{eq:appendix_tsionas_density}, is a sampling density, used for Bayesian analysis using Gibbs sampling. However, to incorporate this density as an error distribution, \eqref{eq:appendix_tsionas_density} should be a proper density function. We therefore prove that the expression in \eqref{Eq:MVG} is indeed a valid probability density function.

\subsection{Derivation and normalization of the corrected multivariate Gamma density}
\label{app:corrected_density}

The representation in \eqref{eq:tsionas_rv} implies that all components $X_i$ share the same latent Gamma random variable $Z$. Conditional on $Z=z$, we have
\[
X_i\mid Z=z=V_i+z,
\qquad i=1,\ldots,d.
\]

Since $V_1,\ldots,V_d$ are mutually independent and independent of $Z$, the components of $\mathbf{X}$ are conditionally independent given $Z$.

Consequently, their conditional joint
density is
\begin{equation}
f_{\mathbf{X}\mid Z}(\mathbf{x}\mid z)
=
\prod_{i=1}^d f_{V_i}(x_i-z),
\qquad 0<z<B.
\label{eq:conditional_mvg_density}
\end{equation}

Integrating the conditional joint density
with respect to the distribution of the
common factor gives
\begin{equation}
g(\mathbf{x})
=
\int_0^B
f_Z(z)
\prod_{i=1}^d f_{V_i}(x_i-z)\,dz,
\label{eq:mvg_marginal_construction}
\end{equation}
where $B=\min_i x_i$.

Substituting the corresponding Gamma
densities into
\eqref{eq:mvg_marginal_construction}
yields
\begin{equation}
\begin{aligned}
g(\mathbf{x})
&=
\frac{\beta^\alpha}{\Gamma(\alpha)}
\prod_{i=1}^d
\frac{\theta_i^{\lambda_i}}
{\Gamma(\lambda_i)}
\times
\int_0^B
\left\{
\prod_{i=1}^d
(x_i-z)^{\lambda_i-1}
e^{-\theta_i(x_i-z)}
\right\}
z^{\alpha-1}e^{-\beta z}\,dz
\end{aligned}
\label{eq:corrected_mvg_density}
\end{equation}

Unlike \eqref{eq:tsionas_rv}, \eqref{eq:corrected_mvg_density} contains a single integral because all
components share the same latent  variable $Z$. The following proposition establishes that this expression defines a valid probability density.

\begin{proposition}
\label{prop:corrected_normalization}
Let $d\geq1$ and suppose that
$\alpha,\beta,\lambda_i,\theta_i>0$
for $i=1,\ldots,d$.
Then the corrected multivariate Gamma
density in
\eqref{eq:corrected_mvg_density}
is nonnegative and satisfies
\[
\int_{(0,\infty)^d}
g(\mathbf{x})\,d\mathbf{x}=1
\]
\end{proposition}

\begin{proof}
Since the Gamma densities are nonnegative, the integrand in \eqref{eq:mvg_marginal_construction} is nonnegative, implying that $g(\mathbf{x})\geq0$.
Using
\eqref{eq:mvg_marginal_construction},
we have
\begin{align}
\int_{(0,\infty)^d}
g(\mathbf{x})\,d\mathbf{x}
&=
\int_{(0,\infty)^d}
\int_0^{\min_i x_i}
f_Z(z)
\prod_{i=1}^d
f_{V_i}(x_i-z)
\,dz\,d\mathbf{x}
\label{eq:mvg_normalization_start}
\end{align}
Because the integrand is nonnegative, Tonelli's theorem permits the order of integration to be interchanged. For a fixed $z>0$, the restriction $0<z<\min_i x_i$ is equivalent to $x_i>z$ for every $i$.
Therefore,
\begin{align}
\int_{(0,\infty)^d}
g(\mathbf{x})\,d\mathbf{x}
&=
\int_0^\infty
f_Z(z)
\left\{
\int_{(z,\infty)^d}
\prod_{i=1}^d
f_{V_i}(x_i-z)
\,d\mathbf{x}
\right\}dz
\label{eq:mvg_normalization_tonelli}
\end{align}
For fixed $z$, make the transformation $v_i=x_i-z$, $i=1,\ldots,d$. The Jacobian determinant, $|\mathcal{J}|=1$ 
and the domain becomes $(0,\infty)^d$.
Thus,
\begin{align}
\int_{(0,\infty)^d}
g(\mathbf{x})\,d\mathbf{x}
&=
\int_0^\infty
f_Z(z)
\left\{
\int_{(0,\infty)^d}
\prod_{i=1}^d
f_{V_i}(v_i)
\,d\mathbf{v}
\right\}dz
\nonumber\\
&=
\int_0^\infty
f_Z(z)
\prod_{i=1}^d
\left\{
\int_0^\infty
f_{V_i}(v_i)\,dv_i
\right\}dz
\nonumber\\
&=
\int_0^\infty f_Z(z)\,dz
\nonumber\\
&=1.
\label{eq:mvg_normalization_result}
\end{align}

Hence,
\eqref{eq:corrected_mvg_density}
defines a properly normalized
multivariate probability density.
\end{proof}

\subsection{Moments and dependence structure
of the corrected multivariate Gamma density}
\label{app:mvg_moments}

We next establish that the corrected
density preserves the theoretical
moments and dependence structure
implied by the common-factor
representation in
\eqref{eq:tsionas_rv}.
The following proposition gives the
joint moment-generating function,
expectation, and covariance.

\begin{proposition}
\label{prop:mvg_moments}
Let $\mathbf{X}=(X_1,\ldots,X_d)^\top$
have the corrected multivariate Gamma
density in
\eqref{eq:corrected_mvg_density}.
Then its joint moment-generating
function is
\begin{equation}
M_{\mathbf{X}}(\mathbf{t})
=
\left(
\frac{\beta}
{\beta-\sum_{i=1}^d t_i}
\right)^\alpha
\prod_{i=1}^d
\left(
\frac{\theta_i}{\theta_i-t_i}
\right)^{\lambda_i}
\label{eq:mvg_mgf}
\end{equation}
for $t_i<\theta_i$,
$i=1,\ldots,d$, and
$\sum_{i=1}^d t_i<\beta$.
Furthermore, for $i,j=1,\ldots,d$,
\begin{equation}
\mathbb{E}(X_i)
=
\frac{\lambda_i}{\theta_i}
+
\frac{\alpha}{\beta},
\label{eq:mvg_mean}
\end{equation}
\begin{equation}
\operatorname{Cov}(X_i,X_j)
=
\frac{\lambda_i}{\theta_i^2}
\mathbf{1}_{\{i=j\}}
+
\frac{\alpha}{\beta^2}
\label{eq:mvg_covariance}
\end{equation}
where $\mathbf{1}_{\{i=j\}}$ is one
when $i=j$ and zero otherwise.
\end{proposition}

\begin{proof}
From the common-factor representation,
\[
X_i=V_i+Z,\qquad i=1,\ldots,d.
\]

For $\mathbf{t}=(t_1,\ldots,t_d)^\top$,
the joint moment-generating function is
\begin{align}
M_{\mathbf{X}}(\mathbf{t})
&=
\mathbb{E}
\left[
\exp\left(
\sum_{i=1}^d t_iX_i
\right)
\right]
\nonumber\\
&=
\mathbb{E}
\left[
\exp\left(
\sum_{i=1}^d t_iV_i
+
Z\sum_{i=1}^d t_i
\right)
\right]
\label{eq:mvg_mgf_start}
\end{align}

By the mutual independence of
$V_1,\ldots,V_d$ and $Z$,
\begin{align}
M_{\mathbf{X}}(\mathbf{t})
&=
\prod_{i=1}^d
\mathbb{E}(e^{t_iV_i})
\,
\mathbb{E}
\left[
e^{Z\sum_{i=1}^d t_i}
\right]
\nonumber\\
&=
\prod_{i=1}^d
\left(
\frac{\theta_i}{\theta_i-t_i}
\right)^{\lambda_i}
\left(
\frac{\beta}
{\beta-\sum_{i=1}^d t_i}
\right)^\alpha
\label{eq:mvg_mgf_derived}
\end{align}

This establishes
\eqref{eq:mvg_mgf}.
Taking the logarithm of the
moment-generating function gives
the cumulant-generating function
\begin{align}
K_{\mathbf{X}}(\mathbf{t})
&=
\log M_{\mathbf{X}}(\mathbf{t})
\nonumber\\
&=
\alpha
\left[
\log\beta-
\log\left(
\beta-\sum_{i=1}^d t_i
\right)
\right]
\nonumber\\
&\quad+
\sum_{i=1}^d
\lambda_i
\left[
\log\theta_i-
\log(\theta_i-t_i)
\right]
\label{eq:mvg_cgf}
\end{align}

Differentiating with respect to $t_i$,
\begin{equation*}
\frac{\partial K_{\mathbf{X}}}
{\partial t_i}
=
\frac{\alpha}
{\beta-\sum_{k=1}^d t_k}
+
\frac{\lambda_i}
{\theta_i-t_i}
\end{equation*}

Evaluating at $\mathbf{t}=\mathbf{0}$
yields
\[
\mathbb{E}(X_i)
=
\left.
\frac{\partial K_{\mathbf{X}}}
{\partial t_i}
\right|_{\mathbf{t}=\mathbf{0}}
=
\frac{\alpha}{\beta}
+
\frac{\lambda_i}{\theta_i}
\]

Similarly, the second-order partial
derivatives are
\begin{equation*}
\frac{\partial^2 K_{\mathbf{X}}}
{\partial t_i\partial t_j}
=
\frac{\alpha}
{\left(
\beta-\sum_{k=1}^d t_k
\right)^2}
+
\mathbf{1}_{\{i=j\}}
\frac{\lambda_i}
{(\theta_i-t_i)^2}
\label{eq:mvg_second_derivative}
\end{equation*}

Evaluating at $\mathbf{t}=\mathbf{0}$
gives
\begin{align}
\operatorname{Cov}(X_i,X_j)
&=
\left.
\frac{\partial^2 K_{\mathbf{X}}}
{\partial t_i\partial t_j}
\right|_{\mathbf{t}=\mathbf{0}}
\nonumber\\
&=
\frac{\alpha}{\beta^2}
+
\frac{\lambda_i}{\theta_i^2}
\mathbf{1}_{\{i=j\}}
\end{align}

Hence, the corrected density has
the stated expectation and covariance
structure.
\end{proof}

Proposition~\ref{prop:mvg_moments}
confirms that the corrected density
preserves the theoretical moments
implied by the original common-factor
Gamma representation. In particular,
the dependence between distinct
components arises from the shared
latent factor $Z$, whose variance
contributes $\alpha/\beta^2$ to every
off-diagonal covariance.
}

\section*{Statements and Declarations}
\noindent\textbf{Funding:}
The authors received no external funding for this research.
\medskip \\
\noindent\textbf{Competing interests:}
The authors declare no conflict of interests.
\medskip \\
\noindent\textbf{Data availability:}
The empirical analysis uses five daily realized-volatility measures constructed
from intraday transaction data for Microsoft Corporation (MSFT) over 2010--2015.
The project repository (\url{https://github.com/Rohan-Chhatre/sparse_logvMEM})
contains a \texttt{data/} directory and selected numerical outputs.
\medskip \\
\noindent\textbf{Code availability:} 
The R code associated with the proposed estimation procedures, simulation study, and CPU/GPU implementations is publicly available at
\url{https://github.com/Rohan-Chhatre/sparse_logvMEM}.
The repository also provides selected outputs of the rolling-window analysis.
\medskip \\
\noindent\textbf{CRediT Author Statement:}
\textbf{RHC}: Conceptualization, Methodology, Software, Formal analysis, Investigation, Validation, Visualization, Writing – original draft, Writing – review \& editing. \textbf{CD}: Conceptualization, Methodology, Writing – original draft, \textbf{SB}: Conceptualization, Methodology, Writing - Review \& Editing, Supervision. \textbf{NR}: Methodology, Conceptualization, Writing - Review \& Editing, Supervision 
\medskip \\
\noindent\textbf{Acknowledgements:} 
{
The authors would like to express thanks to the UConn Storrs HPC support staff for their efforts in maintaining the computational resources utilized in this study.}
\bibliographystyle{apalike} 
\bibliography{biblio-HFT-survey}

@article{Barndorff2004,
  title = {Econometric Analysis of Realized Covariation: High Frequency Based Covariance,  Regression,  and Correlation in Financial Economics},
  volume = {72},
  ISSN = {1468-0262},
  url = {http://dx.doi.org/10.1111/j.1468-0262.2004.00515.x},
  DOI = {10.1111/j.1468-0262.2004.00515.x},
  number = {3},
  journal = {Econometrica},
  publisher = {The Econometric Society},
  author = {Barndorff-Nielsen,  Ole E. and Shephard,  Neil},
  year = {2004},
  month = May,
  pages = {885–925}
}

@article{wright2015coordinate,
  title={Coordinate descent algorithms},
  author={Wright, Stephen J},
  journal={Mathematical Programming},
  volume={151},
  number={1},
  pages={3--34},
  year={2015},
  publisher={Springer}
}

@article{tseng2001convergence,
  title={Convergence of a Block Coordinate Descent Method for Nondifferentiable Minimization},
  author={Tseng, Paul},
  journal={Journal of Optimization Theory and Applications},
  volume={109},
  number={3},
  pages={475--494},
  year={2001}
}

@book{bertsekas1999nonlinear,
  title={Nonlinear Programming},
  author={Bertsekas, Dimitri P.},
  year={1999},
  publisher={Athena Scientific}
}

@techreport{powell2009bobyqa,
  title={The BOBYQA Algorithm for Bound Constrained Optimization Without Derivatives},
  author={Powell, M. J. D.},
  institution={University of Cambridge},
  year={2009}
}

@manual{johnson2024nloptr,
  title={nloptr: R Interface to NLopt},
  author={Johnson, Steven G. and others},
  year={2024}
}

@article{taylor2017logarithmic,
  title={The logarithmic vector multiplicative error model: an application to high frequency NYSE stock data},
  author={Taylor, Nick and Xu, Yongdeng},
  journal={Quantitative Finance},
  volume={17},
  number={7},
  pages={1021--1035},
  year={2017},
  publisher={Taylor \& Francis}
}

@article{nicholson2020high,
  title={High Dimensional Forecasting via Interpretable Vector Autoregression.},
  author={Nicholson, William B and Wilms, Ines and Bien, Jacob and Matteson, David S},
  journal={J. Mach. Learn. Res.},
  volume={21},
  number={166},
  pages={1--52},
  year={2020}
}

@article{Basu2015b,
  author  = {Basu,  Sumanta and Ma, Jing and Michailidis, George},
  title   = {Network Granger Causality with Inherent Grouping Structure},
  journal = {Journal of Machine Learning Research},
  year    = {2015},
  volume  = {17},
  number  = {166},
  pages   = {417-453},
  url     = {https://jmlr.org/papers/v16/basu15a.html}
}

@article{Basu2015a,
  title = {Regularized estimation in sparse high-dimensional time series models},
  volume = {43},
  ISSN = {2168-8966},
  url = {http://dx.doi.org/10.1214/15-AOS1315},
  DOI = {10.1214/15-aos1315},
  number = {4},
  journal = {The Annals of Statistics},
  publisher = {Institute of Mathematical Statistics},
  author = {Basu,  Sumanta and Michailidis,  George},
  year = {2015},
  month = Aug,
  pages = {1535–1567}
}

@article{cattivelli2020adaptive,
 title = {Adaptive Lasso for vector Multiplicative Error Models},
  volume = {20},
  ISSN = {1469-7696},
  url = {http://dx.doi.org/10.1080/14697688.2019.1651451},
  DOI = {10.1080/14697688.2019.1651451},
  number = {2},
  journal = {Quantitative Finance},
  publisher = {Informa UK Limited},
  author = {Cattivelli,  Luca and Gallo,  Giampiero M.},
  year = {2019},
  month = Aug,
  pages = {255–274}
}

@article{mathai1991multivariate,
  title={On a multivariate gamma},
  author={Mathai, Arak M and Moschopoulos, Panagis G},
  journal={Journal of Multivariate Analysis},
  volume={39},
  number={1},
  pages={135--153},
  year={1991},
  publisher={Elsevier}
}

@article{tsionas2004bayesian,
  title={Bayesian inference for multivariate gamma distributions},
  author={Tsionas, Efthymios G},
  journal={Statistics and Computing},
  volume={14},
  number={3},
  pages={223--233},
  year={2004},
  publisher={Springer}
}

@article{cipollini2013semiparametric,
  title={Semiparametric vector {MEM}},
  author={Cipollini, Fabrizio and Engle, Robert F and Gallo, Giampiero M},
  journal={Journal of Applied Econometrics},
  volume={28},
  number={7},
  pages={1067--1086},
  year={2013},
  publisher={Wiley Online Library}
}

@article{breheny2015group,
  title={Group descent algorithms for nonconvex penalized linear and logistic regression models with grouped predictors},
  author={Breheny, Patrick and Huang, Jian},
  journal={Statistics and Computing},
  volume={25},
  number={2},
  pages={173--187},
  year={2015},
  publisher={Springer}
}

@article{zhang2010nearly,
  title={Nearly unbiased variable selection under minimax concave penalty},
  author={Zhang, Cun-Hui},
  journal={The Annals of Statistics},
  volume={38},
  number={2},
  pages={894--942},
  year={2010},
  publisher={Institute of Mathematical Statistics}
}

@article{fan2001variable,
  title={Variable selection via nonconcave penalized likelihood and its oracle properties},
  author={Fan, Jianqing and Li, Runze},
  journal={Journal of the American Statistical Association},
  volume={96},
  number={456},
  pages={1348--1360},
  year={2001},
  publisher={Taylor \& Francis}
}

@misc{cipollini2006vector,
  title={Vector multiplicative error models: representation and inference},
  author={Cipollini, Fabrizio and Engle III, Robert F and Gallo, Giampiero M},
  year={2006},
  publisher={National Bureau of Economic Research Cambridge, Mass., USA}
}

@book{hautsch2011econometrics,
  title={Econometrics of financial high-frequency data},
  author={Hautsch, Nikolaus},
  year={2011},
  publisher={Springer Science \& Business Media}
}

@article{chou2005forecasting,
  title={Forecasting financial volatilities with extreme values: the conditional autoregressive range ({CARR}) model},
  author={Chou, Ray Yeutien},
  journal={Journal of Money, Credit and Banking},
  pages={561--582},
  year={2005},
  publisher={JSTOR}
}

@article{manganelli2005duration,
  title={Duration, volume and volatility impact of trades},
  author={Manganelli, Simone},
  journal={Journal of Financial Markets},
  volume={8},
  number={4},
  pages={377--399},
  year={2005},
  publisher={Elsevier}
}

@article{donelli2021bayesian,
  title={A {B}ayesian semiparametric vector Multiplicative Error Model},
  author={Donelli, Nicola and Peluso, Stefano and Mira, Antonietta},
  journal={Computational Statistics \& Data Analysis},
  volume={161},
  pages={107242},
  year={2021},
  publisher={Elsevier}
}

@article{cipollini2017copula,
  title={Copula--based v{MEM} specifications versus alternatives: the case of trading activity},
  author={Cipollini, Fabrizio and Engle, Robert F and Gallo, Giampiero M},
  journal={Econometrics},
  volume={5},
  number={2},
  pages={16},
  year={2017},
  publisher={MDPI}
}

@article{cipollini2022multiplicative,
  title={Multiplicative Error Models: 20 years on},
  author={Cipollini, Fabrizio and Gallo, Giampiero M},
  journal={Econometrics and Statistics},
  year={2022},
  publisher={Elsevier}
}

@article{yan2017hierarchical,
  title={Hierarchical sparse modeling: A choice of two group lasso formulations},
  author={Yan, Xiaohan and Bien, Jacob},
  journal={Statistical Science},
  volume={32},
  number={4},
  pages={531--560},
  year={2017},
  publisher={Institute of Mathematical Statistics}
}

@article{zhao2009composite,
  title={The composite absolute penalties family for grouped and hierarchical variable selection},
  author={Zhao, Peng and Rocha, Guilherme and Yu, Bin},
  journal={The Annals of Statistics},
  volume={37},
  number={6A},
  pages={3468--3497},
  year={2009},
  publisher={Institute of Mathematical Statistics}
}

@article{lim2015learning,
  title={Learning interactions via hierarchical group-lasso regularization},
  author={Lim, Michael and Hastie, Trevor},
  journal={Journal of Computational and Graphical Statistics},
  volume={24},
  number={3},
  pages={627--654},
  year={2015},
  publisher={Taylor \& Francis}
}

@article{bauwens2000logarithmic,
  title={The logarithmic ACD model: an application to the bid-ask quote process of three NYSE stocks},
  author={Bauwens, Luc and Giot, Pierre},
  journal={Annales d'Economie et de Statistique},
  pages={117--149},
  year={2000},
  publisher={JSTOR}
}

@book{lutkepohl2005new,
  title={New introduction to Multiple Time Series Analysis},
  author={L{\"u}tkepohl, Helmut},
  year={2005},
  publisher={Springer Science \& Business Media}
}

@article{wilms2021sparse,
  title={Sparse identification and estimation of large-scale vector autoregressive moving averages},
  author={Wilms, Ines and Basu, Sumanta and Bien, Jacob and Matteson, David S},
  journal={Journal of the American Statistical Association},
  pages={1--12},
  year={2021},
  publisher={Taylor \& Francis}
}

@article{parkinson1980extreme,
  title={The extreme value method for estimating the variance of the rate of return},
  author={Parkinson, Michael},
  journal={Journal of Business},
  pages={61--65},
  year={1980},
  volume={53},
  number={1},
  publisher={JSTOR}
}

@article{liu2015does,
  title={Does anything beat 5-minute {RV}? A comparison of realized measures across multiple asset classes},
  author={Liu, Lily Y and Patton, Andrew J and Sheppard, Kevin},
  journal={Journal of Econometrics},
  volume={187},
  number={1},
  pages={293--311},
  year={2015},
  publisher={Elsevier}
}

@article{andersen2006volatility,
  title={Volatility and correlation forecasting},
  author={Andersen, Torben G and Bollerslev, Tim and Christoffersen, Peter F and Diebold, Francis X},
  journal={Handbook of Economic Forecasting},
  volume={1},
  pages={777--878},
  year={2006},
  publisher={Elsevier}
}

@article{andersen2012jump,
  title={Jump-robust volatility estimation using nearest neighbor truncation},
  author={Andersen, Torben G and Dobrev, Dobrislav and Schaumburg, Ernst},
  journal={Journal of Econometrics},
  volume={169},
  number={1},
  pages={75--93},
  year={2012},
  publisher={Elsevier}
}

@article{barndorff2009realized,
title={Realized kernels in practice: trades and quotes},
  author={Barndorff-Nielsen, Ole E and Hansen, P Reinhard and Lunde, Asger and Shephard, Neil},
  year={2009},
   journal={The Econometrics Journal},
   volume ={12},
   number ={3},
   pages ={1--32},
  publisher={Oxford University Press Oxford, UK}
}

@article{engle2002new,
  title={New frontiers for {ARCH} models},
  author={Engle, Robert},
  journal={Journal of Applied Econometrics},
  volume={17},
  number={5},
  pages={425--446},
  year={2002},
  publisher={Wiley Online Library}
}

@article{engle1998autoregressive,
  title={Autoregressive conditional duration: a new model for irregularly spaced transaction data},
  author={Engle, Robert F and Russell, Jeffrey R},
  journal={Econometrica},
  pages={1127--1162},
  year={1998},
  publisher={JSTOR}
}

@article{ahoniemi2009joint,
  title={Joint modeling of call and put implied volatility},
  author={Ahoniemi, Katja and Lanne, Markku},
  journal={International Journal of Forecasting},
  volume={25},
  number={2},
  pages={239--258},
  year={2009},
  publisher={Elsevier}
}

@article{he2004extended,
  title={An extended constant conditional correlation {GARCH} model and its fourth-moment structure},
  author={He, Changli and Ter{\"a}svirta, Timo},
  journal={Econometric Theory},
  volume={20},
  number={5},
  pages={904--926},
  year={2004},
  publisher={Cambridge University Press}
}

@Manual{belfrage2015acdm,
    title = {ACDm: Tools for Autoregressive Conditional Duration Models},
    author = {Markus Belfrage},
    year = {2024},
    note = {R package version 1.0.4.3},
    url = {https://CRAN.R-project.org/package=ACDm},
  }

@article{nicholson2017varx,
  title={{VARX-L}: Structured regularization for large vector autoregressions with exogenous variables},
  author={Nicholson, William B and Matteson, David S and Bien, Jacob},
  journal={International Journal of Forecasting},
  volume={33},
  number={3},
  pages={627--651},
  year={2017},
  publisher={Elsevier}
}
\end{document}